\documentclass[preprint]{aastex}
\newcommand{\btex}{0}
\newcommand{\aastex}{1}
\newcommand{\preprint}{1}

\if1\preprint\usepackage{graphicx,psfrag}\fi
\usepackage{natbib}
\if1\aastex
\shorttitle{Bayesian Spectral Analysis}
\shortauthors{van Dyk, Connors, Kashyap, and Siemiginowska}
\else

\fi

\begin{document}

\if1\btex\bibliographystyle{/home/vandyk/TeXfiles/BibTeX/natbib}\fi

\if0\aastex
    \newcommand\apj{ApJ}
    \newcommand\aj{AJ}
    \newcommand\apjl{ApJ}
    \newcommand\aap{A\&A}
    \newcommand\nat{Nature}
\fi

\newcommand\dist{\buildrel\rm d\over\sim}
\newcommand\bY{{\bf Y}} 
\newcommand\bI{{\bf I}} 
\newcommand\bX{{\bf X}} 
\newcommand\bM{{\bf M}} 
\newcommand\bd{{\bf d}} 
\newcommand\bp{{\bf p}} 
\newcommand\bin{^{\rm bin}}
\newcommand\obs{^{\rm obs}}
\newcommand\aug{^{\rm aug}}
\newcommand\mis{^{\rm mis}}
\newcommand\Ya{\dot {\bf Y}}
\newcommand\Yb{\ddot {\bf Y}}
\newcommand\mYa{\dot Y}
\newcommand\mYb{\ddot Y}
\if0\aastex 
\newcommand\Yc{{\bf Y}\kern -.65pc{}^{^{\textstyle\cdot\kern -3pt\cdot\kern -3pt\cdot}}
          \hskip -0.75pc\phantom{Y}}
\newcommand\mYc{Y\kern -.65pc{}^{^{\textstyle\cdot\kern -3pt\cdot\kern -3pt\cdot}}
          \hskip -0.75pc\phantom{Y}}
\else
\if0\preprint 
\newcommand\Yc{{\bf Y}\kern -.82pc{}^{^{\textstyle\cdot\kern -3pt\cdot\kern -3pt\cdot}}
          \hskip -0.75pc\phantom{Y}}
\newcommand\mYc{Y\kern -.75pc{}^{^{\textstyle\cdot\kern -3pt\cdot\kern -3pt\cdot}}
          \hskip -0.75pc\phantom{Y}}
\else 
\newcommand\Yc{{\bf Y}\kern -.72pc{}^{^{\textstyle\cdot\kern -3pt\cdot\kern -3pt\cdot}}
          \hskip -0.75pc\phantom{Y}}
\newcommand\mYc{Y\kern -.7pc{}^{^{\textstyle\cdot\kern -3pt\cdot\kern -3pt\cdot}}
          \hskip -0.75pc\phantom{Y}}
\fi
\fi
\newcommand\eY{{\cal Y}}
\newcommand\eYa{\dot{\cal Y}}
\newcommand\eYb{\ddot{\cal Y}}
\newcommand\cJ{{\cal J}}
\newcommand\cK{{\cal K}}
\newcommand\cL{{\cal L}}
\newcommand\cS{{\cal S}}

\newcommand\lamc{\lambda^L}
\newcommand\alc{\alpha}
\newcommand\itoN{^N_{i=1}}
\newcommand\iton{^n_{i=1}}
\newcommand\itoNs{^{N^s}_{i=1}}
\newcommand\jtoJ{^J_{j=1}}
\renewcommand\th{\theta}
\newcommand\ind{\stackrel{\rm indep.}{\sim}}
\newcommand\updot{^+}
\newcommand\downdot{_+}
\renewcommand\r{\right}
\renewcommand\l{\left}
\newcommand\Var{{\rm Var}}
\newcommand\E{{\rm E}}
\newcommand\N{{\rm N}}
\newcommand\cur{_{[t]}}
\newcommand\nex{_{[t+1]}}
\if1\aastex\renewcommand\pre{_{[t-1]}}
\else\newcommand\pre{_{[t-1]}}\fi
\newcommand{\btheta}{\mbox{\boldmath{$\theta$}}}
\newcommand{\bphi}{\mbox{\boldmath{$\phi$}}}
\newcommand{\blambda}{\mbox{\boldmath{$\lambda$}}}

\newcommand\spacingset[1]{\renewcommand{\baselinestretch}%
{#1}\small\normalsize}

\spacingset{1.2}

\title{\bf Analysis of Energy Spectra with Low Photon Counts via
  Bayesian Posterior Simulation}

\if0\aastex
\author{David A. van Dyk
  \thanks{
    The authors gratefully acknowledge
    funding for this project partially provided by NSF grant
    DMS-97-05157, the Chandra X-ray
    Observatory, and by NASA grant NAS8-39073. They also thank Rostislav
    Protassov  and
    C.J. Shen for their work on the programming of the various algorithms,
    Richard Edgar and Paul Gorenstein for helpful comments on an earlier 
    draft, and the referee, Jeff Scargle, whose careful reading and detailed
    suggestions have greatly improved the paper.}\\
  Department of Statistics, Harvard University
\and
Alanna Connors\\
Eurika Scientific
\and
Vinay L. Kashyap and Aneta Siemiginowska \\
Harvard-Smithsonian Center for Astrophysics}

\maketitle
\fi

\if1\aastex
\author{David A. van Dyk}
\affil{Department of Statistics, Harvard University}
\affil{One Oxford Street, Cambridge, MA 02138}
\email{vandyk@stat.harvard.edu}
\and
\author{Alanna Connors\altaffilmark{1}}
\affil{Department of Astronomy, Wellesley College}
\affil{Whitin Observatory, Wellesley College, Wellesley MA 02481}
\email{connors@frances.astro.wellesley.edu}
\altaffiltext{1}{Alanna Connors is currently affiliated
with Eureka Scientific, 2452 Delmer Street Suite 100, Oakland CA 94602-3017.}
\and
\author{Vinay L. Kashyap and Aneta Siemiginowska}
\affil{Harvard-Smithsonian Center for Astrophysics}
\affil{60 Garden Street, Cambridge, MA 02138}
\email{kashyap@head-cfa.harvard.edu aneta@head-cfa.harvard.edu }
\fi

\medskip


\begin{abstract}
Over the past ten years Bayesian methods have rapidly grown 
more popular in many scientific disciplines as several
computationally intensive statistical algorithms 
have become feasible with increased computer power. 
In this paper, we begin with a general description 
of the Bayesian paradigm for
statistical inference and the various state-of-the-art model fitting
techniques that we employ (e.g., the Gibbs sampler and the
Metropolis-Hastings algorithm). These algorithms are very flexible
and can be used to fit models that account for the highly hierarchical
structure inherent in the collection of high-quality 
spectra and thus can keep pace
with the accelerating progress of new space telescope designs.
The methods we develop, which will soon be available in the Chandra
Interactive Analysis of Observations (CIAO) software, explicitly model
photon arrivals as a Poisson 
process and, thus, have no difficulty with  high resolution low count
X-ray and $\gamma$-ray data. We expect these methods to be
useful not only for the recently launched Chandra X-ray observatory
and XMM but also new generation telescopes such 
as Constellation X, GLAST, etc.  
In the context of two
examples (Quasar S5 0014+813  and Hybrid-Chromosphere 
Supergiant Star $\alpha$~TrA) we illustrate a new highly structured model 
and how Bayesian
posterior sampling can be used to compute estimates, error 
bars, and credible intervals for the various model parameters. 
Application of our method to the high-energy tail of the ASCA spectrum
of $\alpha$~TrA confirms that even at a quiescent state, the coronal
plasma on this hybrid-chromosphere star is indeed at high temperatures
($>10$ MK) that normally characterize flaring plasma on the Sun.  We
are also able to constrain the coronal metallicity, and find that though
it is subject to large uncertainties, it is consistent with the photospheric
measurements.

\end{abstract}

\if1\aastex\fi

\section{Introduction}
The ever increasing power and sophistication of today's high energy
instruments is opening a new realm of high quality data that 
is quickly pushing
beyond the capabilities of the ``classical'' data-analysis methods
in common use.  In this paper, we present an innovative implementation of 
state-of-the-art statistical methods for fitting high resolution 
spectra from the Chandra X-ray Observatory. 
The common ``folk-wisdom'' of how to bin data, 
subtract background counts, propagate errors, and, for example,
estimate the significance of a spectral line profile are unreliable
and can lead to unacceptable results
\citep[see][for discussion]{lore:93, nous:93, feig:babu:97, siem:etal:97, 
zimm:97}. For example, binning data
sacrifices the resolution of the instrument, subtracting background
can lead to negative counts with unpredictable results, and statistical
black boxes such as the $\chi^2$ and Cash statistics 
\citep{lamp:marg:bowy:76,cash:79} although often useful, may not be 
equipped to answer standard questions \citep[e.g.,][]{vand:prot:00}.  Some
authors have suggested solutions to such problems which
involve ad hoc adaptations of commonly used methods 
\citep[e.g.,][]{gehr:86, coll:etal:87, migh:99}.   
Unfortunately, when such solutions are not rooted in a
theoretical framework, they have no justification beyond problems
which are more-or-less the same as the simulation studies which
justify them and we are often forced into additional ad hoc 
adaptations.  This approach is difficult to justify in light of
modern statistical methods which address reasonable model assumptions
directly.  Thus, in recent years, astrophysicists have increasingly turned to
likelihood based
\citep[e.g.,][]{lucy:74, cash:79, schm:85, scio:mice:92}
and Bayesian methods
\citep[e.g.,][]{bija:71, rich:72, greg:lore:92, lore:93, conn:97, 
siem:97, kash:drak:98, free:etal:99}.
The primary purpose of this paper is to illustrate how
Bayesian methods can provide practical answers to outstanding real
problems which standard methods are not able to handle. The methods
described here are equipped with readily available parameter estimates, 
credible intervals, error bars, model checking techniques, 
methods for combining information from multiple sources, etc.---all 
within a flexible theoretical framework and without reliance on 
asymptotic Gaussian approximations.

We illustrate Bayesian data analysis via two detailed examples.
The analysis of Quasar S5 0014+813 offers a straight forward introduction
to our methods
and the extremely low count Hyprid-Chromosphere 
Supergiant Star $\alpha$~TrA observation shows how we tackle a previously
intractable analysis. Together, these
examples demonstrate the power of Bayesian methods to handle
highly structured models designed to reflect the structure in
both the source spectrum and the
data collection process. Our methods avoid the binning of counts
and thus the sacrificing of high-resolution information required 
by standard data analysis methods. The analysis
of Quasar S5 0014+813 is consistent with the available standard analysis 
which relies on extra binning and the removal of the high-energy low-count tail.
 
We emphasize that we {model} not
only the source spectrum, but also other stochastic components
of data collection and the instrument such as background
contamination
and instrument response. Generally we refer to our stochastic
representation of the entire process as the (statistical) {\it model}.
For clarity we refer to the spectral or physical model as the {\it source
model} and to the model for the observed (PHA\footnote{Pulse Height
Amplitude, originally in proportional counters, the number of electrons 
produced by a photon, hence the amplitude of the current pulse registered 
by the detector electronics. The term now refers to the measure of the
energy deposited on the detector (as opposed to the true energy).}) 
counts as the {\it
observed-data model}. In our detailed example, we develop a model 
and algorithms for
spectral analysis of high-energy (or other) data using
a Poisson\footnote{Recall, a random variable 
$X$ is said to follow a Poisson distribution with parameter or
intensity $\lambda$ if $\Pr(X=x)=e^{-\lambda}\lambda^x
/x!$.  In this case $\E(X)=\lambda$ and we often write
$X\dist{\rm Poisson}(\lambda)$ (read as $X$ is distributed as
Poisson with intensity $\lambda$). This representation conditions on 
the intensity parameter, $\lambda$, which in turn may vary.}  
process for photon arrivals. We allow for:
stochastic instrument response via a photon redistribution matrix;
the absorption of photons; the effective area\footnote{{ The effective
area of the telescope is the fraction of the true geometric area that the 
telescope presents to sky. This varies with energy.}}
of the telescope; and
background contamination of the source.  In particular, we model 
information on background emissions as the realization of a second Poisson
process \citep[cf.][]{lore:93}, thereby eliminating the need to directly 
subtract off the
background counts and the rather embarrassing resulting problem of
negative photon counts. The source energy spectrum is modeled as a mixture of
several (Gaussian) line profiles and a {\it Generalized Linear 
Model (GLM)}\footnote{In a GLM we assume a transformation (e.g., log) of 
the model is  linear in a set 
of independent variables. We emphasize that this is not equivalent to
transforming the data and proceeding with linear regression. A
generalized linear model utilizes the likelihood of the assumed
model which may not be Gaussian. (E.g., the assumed model may be Poisson.)
See Section~3.3 for details.
}
\citep[e.g.,][]{mccu:neld:89}   
which accounts for the continuum. GLMs have
become the standard statistical method for incorporating information 
contained in independent variables (as in regression) into many 
non-Gaussian models and are thus an obvious
but innovative choice in this setting.

In addition to several {\it Markov chain Monte Carlo (MCMC)} algorithms,
we describe and use {\it data augmentation}, an important statistical
method for Bayesian (and other) analysis.  Data augmentation is
an elegant computational construct allowing us to take
advantage of the fact that if it were possible to collect additional
data, statistical analysis would be greatly simplified.  This
is true regardless of {\it why} the so-called ``missing data''
are not observed.
For example, if we were able to record the counts due to background
contamination in addition to total counts in each bin, it would, 
of course, be a trivial task to account for the background.  
There is a large class of powerful statistical methods 
designed for ``missing data''
problems. With the insight that  ``true'' values of quantities
recorded with measurement error can be regarded as ``missing data,''
these methods can usefully be applied to almost any astrophysical
problem.  In particular, we can
treat the true image (before instrument response), the absorbed 
photon counts,
and unbinned energies as ``missing data'' to account for instrument
response, absorption, and binning, respectively.

This introduction fore shadows the tone of the paper: in the process
of developing new Bayesian methods, we describe and utilize
state-of-the-art statistical reasoning, methods, and algorithms,
whereby we explore a larger statistical framework for our problem
of interest. Although we introduce many new tools and use terminology
that may be unfamiliar, we endeavor to write in a manner accessible to 
astrophysicists and believe the resulting methods justify the
required interdisciplinary work.   Table~\ref{tbl:index} indexes
terminology used in the paper that may be unfamiliar to some readers.
To aid in the translation from standard statistical notation to
standard astrophysical usage, many equations have been written
according to both standards when notation is introduced.
One notational convention is worthy on mention; we use superscripts
to identify model components (e.g., background or absorption).

The paper is organized as follows.  After a brief overview
of the fundamentals of Bayesian analysis in
Section~2, we lay out our hierarchical statistical 
model\footnote{A hierarchical (statistical) 
model is formulated in terms of unobserved quantities, which are 
themselves statistically modeled. For example, we may assume photons 
first arrive at a detector according to a Poisson process and them are 
randomly redistributed according to a photon redistribution matrix. A 
hierarchal model separates these two random process into two levels 
of a structured model.} which summarizes the
photon collection process and parameterizes many
relevant aspects of the energy spectrum
in Section~3. Two
examples which aim to illustrate a typical data analysis and
the advantages of Bayesian methods in this setting are 
given in Section~4. Section~5 contains
brief concluding remarks. Finally, in
two appendices, we outline such important general 
MCMC methods as data augmentation,
the Gibbs sampler, the Metropolis-Hastings algorithm, and 
judging convergence using multiple chains. We
also describe in detail how we use these 
algorithms to fit the hierarchical source model of Section~3.

\section{Bayesian Analysis}
In this section we outline several important methodological and
computational issues involved with Bayesian analysis using
a simple model that accounts for background in a simplified 
Poisson process to motivate and illustrate 
ideas.  Our introduction is brief and we 
encourage interested readers to consult one of the several
high-quality recent texts on the subject such as
\citet*{gelm:carl:ster:rubi:95},  
\citet{carl:loui:96}, \citet*{gilk:rich:spie:96},
and \citet{sivi:96}.  In Section~3,
we show how the ideas developed here can be used for a detailed
spectral analysis.

\subsection{Prior, Sampling, and Posterior Distributions}

Bayesian probability analysis is fundamentally based on one simple
result known as Bayes's Theorem which allows us to update a 
probability distribution based on new data or other information.  
In particular, knowledge about a (vector) model 
parameter, $\btheta$, 
is summarized by a probability distribution, 
$p(\btheta)$, such that 
\ $\Pr(\btheta \in R)= \int_R p(\btheta)d \btheta$ \ 
for any region\footnote{The notation
$\Pr(\btheta\in R)$ represents the probability that $\btheta$ is in the region
$R$ and is computed as the integral of the probability distribution
of $\btheta$ over $R$.} $R$.  Bayes's Theorem states
\begin{equation} 
p(\btheta|\bY,\bI) = {{p(\bY|\btheta,\bI) p(\btheta|\bI)} \over{p(\bY|\bI)}} \ ,
\label{eq:bayesthm}
\end{equation}
where $\bY$ are the observed data or other new information pertaining to 
$\btheta$ and $\bI$ represents any initial information known before $\bY$
is observed.  Here, $p(\btheta|\bI)$ represents our knowledge prior to observing
$\bY$ and is called the {\it prior distribution}.  The {\it sampling 
distribution} or {\it likelihood}, $p(\bY|\btheta,\bI)$ represents the
likelihood of the 
data given the model parameters, and $p(\btheta|\bY,\bI)$ represents our
updated knowledge regarding $\btheta$ after observing $\bY$ and
is called the {\it posterior distribution}.  Finally, $p(\bY|\bI)$ represents
the unconditional distribution of $\bY$ and acts as the normalizing 
constant for
$p(\btheta|\bY,\bI)$. The functional form of Bayes's Theorem describes how our
prior knowledge should be updated in light of information contained in
the data. The likelihood or sampling distribution is the basis 
for many standard statistical techniques, while the prior and
posterior distributions are specific to Bayesian analysis.

To illustrate Bayes's Theorem, suppose we have observed counts, $Y$,
contaminated with background in a (source) exposure and have observed 
a second exposure of pure background. Throughout this section, we assume 
the source exposure is $\tau^S$ minutes and the pure background exposure 
is $\tau^B$ minutes with both exposures using the same area of the detector.
(We generally use superscripts to represent photon ``sources,'' e.g.,
background or source. Occasionally we use superscripts for powers;
for clarity we place powers outside parentheses.)
To model the source exposure, we assume $Y$ follows a Poisson 
distribution with intensity $\lambda^B
+\lambda^S$, where $\lambda^B$ and $\lambda^S$ represent the 
expected counts during the source exposure 
due to background and source
respectively.  Thus, the likelihood is
\begin{equation}
p(Y|\lambda^B,\lambda^S,\bI) =
e^{-(\lambda^B+\lambda^S)}(\lambda^B+\lambda^S)^Y\big/Y! \ {\rm for} \
Y=0,1,2,\ldots 
\label{eq:like_examp}
\end{equation}
We wish to estimate $\lambda^S$ and treat $\lambda^B$ as
a {\it nuisance} parameter, a parameter that is of little interest,
but must be included in the model. As is detailed below,
an important advantage of Bayesian methods is their ability to
handle nuisance parameters by computing the {\it marginal} posterior
distribution of the parameters of interest. The name ``marginal''
distribution originates with two-way tables of counts where the table
margins sum over one of the variables to give the distribution of the
other variable alone (e.g., its marginal distribution).
Likewise the marginal distribution of the parameter of
interest is computed by integrating over (i.e., averaging
over) the nuisance parameter.



At this point, we specify a prior distribution which allows us to
include {\it a priori} knowledge (e.g., ``allowed parameter ranges'')
from other experiments or other scientific
information. One of the primary advantages of Bayesian analysis is
a well-defined mechanism for the inclusion of information outside the
current data set. In the absence of prior information, we use diffuse or
so-called {\it non-informative} priors, which are ordinarily flat 
and have minimal influence
on the final analysis. The prior distribution itself may be
conveniently parameterized using a set of 
{\it hyperparameters} that can be varied to represent the
researcher's knowledge about the value of the model parameters and the 
degree of certainty of this knowledge. For example, we
use the $\gamma$ distribution\footnote{The $\gamma$  $(\alpha, \beta)$
  distribution is a continuous distribution on the positive real
  line with probability density function $p(Y)=\beta^\alpha Y^{\alpha-1}
  e^{-\beta Y}/\Gamma(\alpha)$, expected value $\alpha/\beta$, and
  variance $\alpha/\beta^2$ for positive $\alpha$ and $\beta$.}
to parameterize prior information for $\lambda^B$  and
$\lambda^S$;  
\begin{equation}
\lambda^B|\bI\dist\gamma(\alpha^B,\beta^B) \ \hbox{and} \
\lambda^S|\bI\dist\gamma(\alpha^S,\beta^S),
\end{equation} 
i.e.,
\begin{equation}
p(\lambda^B|\bI) = (\beta^B)^{\alpha^B} (\lambda^B)^{\alpha^B-1}
  e^{-\beta^B \lambda^B}/\Gamma(\alpha^B) \ \hbox{and} \
p(\lambda^S|\bI) = (\beta^S)^{\alpha^S} (\lambda^S)^{\alpha^S-1}
  e^{-\beta^S \lambda^S}/\Gamma(\alpha^S),
\end{equation} 
where the notation $\dist$ is read ``follows the distribution''
and $\lambda^B$ and
$\lambda^S$ are assumed {\it a priori} independent. The $\gamma$ prior on 
$\lambda^B$ is mathematically equivalent to a Poisson likelihood
resulting from a count equal to  $\alpha^B-1$ obtained with an
exposure of $\beta^B$ times that of the source exposure.  
(By mathematically equivalent, we mean the prior on $\lambda^B$ is
proportional to a Poisson likelihood as a function of $\lambda^B$.)
This leads to
a natural choice of $p(\lambda^B|\bI)=\gamma$ $(\alpha^B=Y^B+1,
\beta^B=\tau^B/\tau^S)$, 
where $Y^B$ are the counts from the background exposure. 
Notice that here (and throughout the paper) we explicitly incorporate
information from the background exposure into the analysis via the 
prior distribution on $\lambda^B$. Thus, the counts from the source 
exposure, $Y$, are treated as the observed data $\bY$ in 
Equation~\ref{eq:bayesthm}. We refer interested readers to 
\citet[][Section~2.7]{gelm:carl:ster:rubi:95} for further discussion
and examples of the $\gamma$ prior distribution with Poisson data.

The equivalence of the $\gamma$ prior for $\lambda^S$ and $\alpha^S-1$
counts during an exposure of $\tau^S\beta^S$ minutes
leads to a natural interpretation of the hyperparameters---for
a relatively non-informative prior we choose $\beta^S$ much
less than one.  To 
illustrate this, we consider two priors, one non-informative and
improper\footnote{An improper distribution is a distribution that is
{\it not} integrable and thus is not technically a distribution. One
should use improper prior distributions only with great care since in 
some cases they lead to improper posterior distributions which are 
uninterpretable.},
$p(\lambda^S|\bI)^{[1]}=\gamma(1,0)\propto 1$,
(illustrated as a dotted line in
Figure~\ref{fig:post_examp}); and one informative, where, let us say
we know from other means that three counts are to be expected in the
same exposure time, hence
$p(\lambda^S|\bI)^{[2]}=\gamma(4,1)$
(solid line in Figure~\ref{fig:post_examp}).
This choice of informative prior is only an example---$\gamma(4,1)$
corresponds to Poisson likelihood resulting from 3 counts with
an exposure time equal to the source exposure (ten minutes).
This is a rather informative prior distribution and is chosen to 
illustrate the effect of very informative prior.
The non-informative prior contains information equivalent to zero 
counts in an exposure of zero seconds.

Using Bayes's Theorem, with $\btheta=(\lambda^B,\lambda^S)$, we can
combine the $\gamma$  priors and the likelihood given in
Equation~\ref{eq:like_examp} to compute the posterior distribution, 
\begin{equation}
p(\lambda^B,\lambda^S|Y,\bI)\propto {e}^{-[\lambda^B(\beta^B+1) +
\lambda^S(\beta^S+1)]} (\lambda^B+\lambda^S)^Y 
(\lambda^B)^{\alpha^B-1}(\lambda^S)^{\alpha^S-1},
\label{eq:post_examp}
\end{equation}
for $\lambda^B\geq 0, \lambda^S\geq 0$.

Nuisance parameters such as $\lambda^B$ pose a monumental difficulty for 
classical statistical analysis which often relies on fixing
nuisance parameters at estimated values. Unfortunately, this
does not account for uncertainty in their estimates
and thus tends to be anti-conservative. (Likewise, floating nuisance 
parameters or ``propagating errors'' when computing error bars is 
essentially a Gaussian assumption, which can lead
to unpredictable results when such an assumption is not justified.)
The Bayesian
solution {\it averages} over the posterior distribution (i.e., the
uncertainty) of the nuisance parameter by computing the {\it marginal} 
(posterior) distribution of the parameters of interest without
Gaussian approximations.
For example, the marginal posterior
distribution of $\lambda^S$ can be computed by (numerical) integration, 
\begin{equation}
p(\lambda^S|Y,\bI)=\int_0^\infty p(\lambda^B,\lambda^S|Y,\bI)d\lambda^B,
\label{eq:marg}
\end{equation}
and is illustrated for the two priors for  $\lambda^S$ in the second
plot of  Figure~\ref{fig:post_examp}, where we assign $Y=1$ and
$Y^B=48$.  In this
example, direct subtraction of background would leave a
``negative count'' of $-1$; no such difficulty occurs with the Bayesian
analysis. (See \citet{lore:93} for another derivation of the
marginal distribution of $\lambda^S$ in this setting.)

Since the source count is small relative to the background count,
we expect a small $\lambda^S$.  Although this is evident in both
posterior distributions in Figure~\ref{fig:post_examp}, the highly 
informative prior distribution centered at $\lambda^S=3$ pulls the 
(solid) posterior towards higher values, thus illustrating the effect 
of an informative prior distribution.  Such sensitivity analyses often
play an important part in Bayesian (or other) data analyses, since
they investigate the sensitivity of the results to the statistical
assumptions (e.g., the choice of prior distribution). 

The posterior distributions
should be interpreted as probability distributions representing the
combined information in the prior and data. For example, a region, $R$, 
such that $\int_R p(\btheta|\bY,\bI)d\btheta = \zeta$ 
is called a $\zeta$-level
credible interval (or credible region if $\btheta$ is multidimensional) 
and we can say $\Pr(\btheta\in R|\bY,\bI)=\zeta$ (e.g., a 67\%, 90\%, or 95\%
credible region). The 90\% credible region for the posterior 
distributions illustrated in Figure~\ref{fig:post_examp} are 
(0.77, 4.24) using the informative prior and (0.04, 3.84) using the 
non-informative prior.
Such probability statements are measures of our 
information regarding the value of the parameter $\btheta$, given the
data and prior information.  This is in contrast to the more traditional 
frequentist definition of probability, which defines
a probability to be the long term frequency of an event generally involving
the data given $\btheta$.  The
posterior distribution is a complete summary of our
information, but
is often summarized by its mean, $\hat{\btheta} = \E(\btheta|\bY,\bI)$ 
and variance, $\Var(\btheta|\bY,\bI)$, or its modes and the 
curvatures at these modes. (The curvatures are most useful when the
posterior is (locally) approximately Gaussian, as is asymptotically 
true under
certain regularity conditions; e.g., \citet{gelm:carl:ster:rubi:95}
chapter 4.)  In the following two sections we  
describe Monte Carlo methods for computing posterior means
and variances and credible regions. To compute posterior modes
(e.g., maximum likelihood estimates) \citet{vand:00astro} develops several 
Expectation Maximization or EM algorithms for use in astrophysical
applications. Posterior modes are often used to compute starting
values for the more robust but computationally demanding
Monte Carl methods; see Appendix~A.2.
The EM algorithm gets its name 
because it iteratively {\it maximizes} the {\it expected} log posterior 
distribution of $\btheta$ given the augmented data.

Although a detailed description is beyond the scope of this paper,
Bayesian methodology is well equipped for problems involving
model selection.  Methods based on Bayes's factors, computing the
relative posterior probabilities of various competing models, and
Bayesian `p-values' are all important and remain areas of active
statistical research  \citep[e.g.,][]{greg:lore:92,vand:prot:00}.

\subsection{Evaluating the Posterior via Monte-Carlo Sampling}

For univariate or small dimensional parameter spaces, we can usually
compute the posterior mean, variance, credible regions, and other summaries 
either analytically or via non-stochastic
numerical methods (e.g., Gaussian quadrature or Laplace's
method). In higher dimensions, however, these methods
can be difficult to implement partially because of the difficulty
in finding the region where the integrand is significantly greater 
than zero. Thus, we often resort to Monte-Carlo
integration. In particular, if we can obtain a sample from
the posterior, $\{\btheta_{[t]},t=1,\ldots,T\}$, Monte Carlo
integration approximates the mean of any function, $g$, of 
the parameter with 
\begin{equation}
\E[g(\btheta)|\bY,\bI]\approx {1\over T}\sum_{t=1}^T g(\btheta_{[t]}),
\label{eq:MCapprox}
\end{equation}
where we assume $\E[g(\btheta)|\bY,\bI]$ exists.
For example, $g(\btheta)=\btheta$ and 
$g(\btheta) = (\btheta - \E(\btheta|\bY,\bI))
(\btheta - \E(\btheta|\bY,\bI))^\prime$ lead to the posterior mean and variance
respectively.  Probabilities, such
as $\zeta=\Pr(\btheta\in R)$ can be computed using $g(\btheta)=I\{\btheta \in R\}$, 
where the function $I$ takes on value 1 if the condition in curly
brackets holds and zero otherwise. Likewise, quantiles of the
distribution can be approximated by the corresponding quantiles of the
posterior sample. In short, a robust data analysis requires only a
sample from the posterior distribution. A general strategy is to
first sample from the posterior distribution and then approximate
various integrals of interest via Monte Carlo.

\subsection{Obtaining a Sample from the Posterior}

The Monte-Carlo approximation methods depend on our
ability to obtain a sample from the posterior distribution. Although in 
some cases the posterior distribution is a well known distribution
and trivial to sample, we must often use sophisticated algorithms 
to obtain a posterior sample. In Appendix A, we
discuss three algorithms which have proven widely applicable
in practice, the {\it Data
Augmentation algorithm} \citep{tann:wong:87}, the {\it Gibbs 
sampler} \citep{metr:rose:rose:tell:tell:53}, and the {\it
Metropolis-Hastings algorithm} \citep{hast:70}. All of these algorithms 
construct a Markov chain with stationary distribution equal to the posterior
distribution \citep[e.g.,][]{gelf:smit:90}; i.e., once the chain has reached 
stationarity, it generates samples which are identically (but not 
independently) distributed according to the posterior distribution. 
These samples can then
be used for Monte Carlo integration as described above; hence
these algorithms are known as Markov chain Monte Carlo or 
MCMC methods \citep[see][for regularity conditions for using
Equation~\ref{eq:MCapprox} with MCMC draws]{tier:96}. From the onset then, it is
clear that three 
important concerns when using MCMC in practice are (1) selecting
starting values for the Markov chain, (2) detecting
convergence of the Markov chain to stationarity, and (3) the effect of the
lack of independence in the posterior draws. These issues are
addressed in Appendix A.2.

The algorithms used to fit the models described in 
Section~3 rely on the method of data 
augmentation. The term `data augmentation' originated with 
computational methods designed to handle missing data, but the method 
is really quite general and often useful when there is no missing data 
per se.  In particular, for Monte Carlo integration we aim to obtain a 
sample from the posterior distribution, $p(\btheta|\bY,\bI)$. In some cases, 
we can
{\it augment} the model to $p(\btheta,\bX|\bY,\bI)$, where $\bX$ may be
missing data or any other unobserved quantity (e.g., counts due
to background). With judicial choice of $\bX$, it may be much easier 
to obtain a sample from $p(\btheta,\bX|\bY,\bI)$ than directly from
$p(\btheta|\bY,\bI)$. Once we have a sample from $p(\btheta,\bX|\bY,\bI)$,
we simply discard the sample of $\bX$ to obtain a sample from 
$p(\btheta|\bY,\bI)$. In Appendix~B.1, we describe how this method can
be used for fitting the models described in Section~3.

\section{Fitting High-Resolution Low-Count Spectra}

\subsection{Model Overview}

In this section, we describe a new class of (statistical) structured
models which simultaneously
describes high-resolution source spectra using Gaussian line
profiles and a GLM for the 
continuum and accounts for background contamination of the image, instrument
response, and absorption. The model may easily be generalized to account
for different line profiles such as the Lorentzian distribution
\citep[e.g.,][]{meng:vand:99}.  The
(statistical) model is designed to 
summarize the distribution of photon energies arriving 
at a detector, which are recorded as counts in a number of energy channels
(e.g., as many as 4096 on Chandra/ACIS).  Newly developed detectors have
much higher resolution than their predecessors, and thus smaller 
expected counts per bin. Independent Poisson distributions are 
therefore more appropriate  for the counts than the commonly 
used Gaussian approximation (e.g., $\chi^2$ fitting).
We parameterize the intensity in bin $j\in\cJ=\{1,\ldots,J\}$,
as the sum of a continuum term and $K$ Gaussian lines.
That is, the expected true counts per bin for a ``perfect'' instrument
with effective area everywhere equal to the maximum possible effective
area\footnote{We use the maximum value of the effective area over the
  spectral energy range of interest in this stage of the analysis. This is
  only a matter of convenience and the full effective area variations
  are included in Equation~\ref{eq:model_obs}.} is

\vbox{

\begin{equation}
\begin{array}{c@{\;=\;\Big[}c@{\;+\;}c@{\Big]}cc}
\hbox{model intensity} & \hbox{continuum} & \hbox{lines} &
\hbox{absorption,} & \hbox{for each energy bin}\\
\multicolumn{4}c{} \\
\lambda_j(\btheta) & dE_j f(\btheta^C,E_j) & 
\displaystyle{\sum_{k=1}^K \tilde\lambda^k p_j (\mu^k,\upsilon^k)}&
u(\btheta^A,E_j), & \hbox{for} \ j\in\cJ,
\end{array}
\label{eq:model_true}
\end{equation}

\if1\aastex\kern -1.0in\else\kern -0.75in\fi

\noindent
or more formally,

\kern 0.4in
}

\noindent
where $dE_j$ is the known width of bin 
$j$, $f(\btheta^C, E_j)$ is the expected number of counts per 
keV per maximum effective area from the continuum
and is a function of the continuum 
parameter, $\btheta^C$, $E_j$ is the known mean energy in bin $j$, 
$\tilde\lambda^k$ are the expected counts per maximum
effective area from line $k$, $p_j(\mu^k,\upsilon^k)$
is the probability that a Gaussian random variable with mean
$\mu^k$ and variance $\upsilon^k$ falls in bin $j$, and
$u(\btheta^A, E_j)$ is the probability that a photon in
bin $j$ is {\it not} absorbed. Specific forms for the continuum and absorption terms are
discussed below in Sections~3.2 and 3.4, respectively. 
  The superscripts on the model parameters ($\btheta$) are mnemonic and represent
absorption (A), background (B), continuum (C), and the lines
$k=1,\ldots, K$.
The collection
of parameters, $\btheta^C,\btheta^k=(\tilde\lambda^k, \mu^k, 
\upsilon^k) \ {\rm for} \ k\in\cK=\{1,\ldots,K\}$, and 
$\btheta^A$ (along with $\btheta^B$ defined below) are represented
by $\btheta$. An artificial example, 
with power law
continuum, two spectral lines, and no absorption appears in the first
plot of Figure~\ref{fig:datagen}.

Since data collection is degraded by effective area, 
instrument response, and background contamination (see Figure~\ref{fig:datagen}), 
we model the observed counts as independent Poisson variables
with intensity

\vbox{

\begin{equation}
\begin{array}{c@{\;=\;}c@{\;+\;}c@{\;\;}c}
\hbox{observed}\atop\hbox{intensity}& 
{\hbox{instrument}\atop\hbox{response}} 
\l({\hbox{model}\atop\hbox{intensity}}
                             *{\hbox{effective}\atop\hbox{area}}\r)&
\hbox{background,} & \hbox{for each}\atop\hbox{energy channel}\\
\multicolumn{4}c{} \\
\xi_l(\btheta) & \displaystyle{\sum_{j\in\cJ}} M_{lj} \lambda_j(\btheta) 
d_j & \lambda^B_l(\btheta^B), & \hbox{for} \ l\in\cL,
\end{array}
\label{eq:model_obs}
\end{equation}

\if1\aastex\kern -0.85in\else\kern -0.65in\fi

\noindent
or more formally,

\kern 0.4in
}

\noindent
$\cL = \{1,\ldots,L\}$
where the $L\times J$ matrix $\bM=\{M_{lj}\}$ represents instrument
response --- a photon arriving in bin $j$ has probability $M_{lj}$
of being detected in observed bin $l$, $\bd=(d_1,\ldots,d_J)$ is the
effective area of bin $j$, normalized so that 
${\rm max}_{j\in\cJ} d_j = 1$, 
and $\lambda^B_l(\btheta^B)$ is the expected
counts due to the background which may be known from calibration in
space or parameterized in terms of $\btheta^B$.
As with $\cJ$, $\cL$ may be any subset of detector bins.
Generally the counts are also degraded by pile up 
\cite[e.g.,][]{knol:89};
see Figure~\ref{fig:datagen}. Here we ignore pile-up
which is justifiable for low intensity or spatially diffuse sources
(see the discussion in Section~5).

In the next several sections, we describe the stochastic models
for each of the sources of photons in turn. This includes both
likelihoods which describe the sampling distribution of the data
(parameterized by $\btheta$) 
and prior distributions which allow us to incorporate
scientific information about the likely parameter values. As
described below, the
prior distributions are parameterized using the hyper-parameter $\bphi$.

\subsection{The Continuum}

The photon counts due to the continuum are modeled via a GLM 
\citep[]{mccu:neld:89}, specifically a log linear model.  
That is, the log expected counts per keV per maximum effective area
are assumed to be a linear function of a set of independent variables, 
$X_j^C$ which in turn are typically functions of $E_j$; hence the
notation, $f(\btheta^C,E_j)$.
In particular we model the counts in bin $j$ due to the continuum,
denoted $Y^C_j$, as 
\begin{equation}
Y^C_j\dist{\rm Poisson}(dE_j f(\btheta^C,E_j)),
\end{equation}
i.e.\footnote{Here and in the remainder of the paper we suppress the 
conditioning on the initial information, $\bI$. That is, it should be 
understood that all distributions implicitly condition on $\bI$.},
\begin{equation}
p(Y^C_j|\btheta^C)=e^{-\lambda^C_j}(\lambda^C_j)^{Y^C_j}/(Y^C_j)! 
\ {\rm with} \ \lambda^C_j=f(\btheta^C,E_j)dE_j.
\end{equation}
Here $\log f(\btheta^C,E_j)=\bX^C_j\btheta^C,$ independently for 
$j\in\cJ,$
with $\btheta^C$ a
$(P^C \times 1)$ vector parameter, $\bX^C_j$ a
$(1\times P^C)$ vector of independent variables, and $P^C$ the
number of parameters in the continuum model.
Note that we are explicitly using a Poisson process for the photon
counts as opposed to an often poor Gaussian approximation.

The flexible 
framework of the GLM allows us to adjust the
expected counts in bin $j$ for any set of independent variables.  
For example, several standard continuum models are easily available.
In particular, a power law model is obtained by setting 
$\bX_j=\l(1, \log(E_j)\r)$ for $j\in\cJ$ so that
\begin{equation}
f(\btheta^C,E_j)=e^{\theta^C_1}E_j^{\theta^C_2} = 
\alpha E^{-\beta}_j \  {\rm for} \  j\in\cJ \ , 
\label{eq:powerlaw}
\end{equation}
where the familiar form of the power law model in the last expression 
is obtained by identifying 
$(\alpha,\beta)$ with $(e^{\theta^C_1},-\theta^C_2)$.  It is easy to
generalize this to handle more complicated models. A
break in the power law (i.e., a change point) can 
be added at $E_{\star}$ by
setting $\bX_j=(1,\log(E_j),\log(E_j/E_*)I\{E_j>E_*\})$,
so that
\begin{equation}
f(\btheta^C,E_j)=
\cases{e^{\theta^C_1}E_j^{\theta^C_2} & for $E_j \leq E_{\star}$\cr
e^{\theta^C_1}E_j^{\theta^C_2+\theta^C_3}E_{\star}^{-\theta^C_3} &
for  $E_j>E_{\star}$} \ {\rm for} \ j\in\cJ \ .
\end{equation}
The factor $E_{\star}^{-\theta^C_3}$ ensures $f(\btheta^C, E_j)$ is
continuous at $E_j = E_{\star}$.  As a final example, we obtain
an exponential continuum representing Bremsstrahlung emission 
by setting $\bX_j^C=(1,-E_j)$ so
\begin{equation}
f(\btheta^C,E_j)=e^{\theta^C_1}e^{-E_j\theta^C_2} = 
{\alpha \over \sqrt{T}} e^{-E_j/kT} \ {\rm for} \
j\in\cJ \ , 
\label{eq:bremss}
\end{equation}
where $(\alpha,T)=(e^{\theta^C_1}\sqrt{k\theta^C_2},1/k\theta^C_2)$.

It is convenient to assume the prior distribution on $\btheta^C$
to is  multivariate
Gaussian with a diagonal variance matrix. That is, $\theta_p^C\dist
\N(\mu_p^C,\upsilon_p^C)$ with $\bphi^C=\{(\mu_p^C,\upsilon_p^C)$,   
for $p=1,\ldots,P^C\}$.
The hyper-parameter, $\bphi^C$, is set by the user where $\mu_p^C$ is a
``best  guess'' of $\theta^C_p$ and $\upsilon^C_p$ is a measure (in
squared standard deviations) of the error of this ``best guess.''
Large values of $\upsilon^C_p$ reflect little prior information for
$\theta_p^C$.

\subsection{Emission Lines}
Lines reflect deviation in the smooth spectrum due to the continuum 
because of  photon emissions from various ions present in the source. In
particular we model the energies of photons due to 
line $k\in\cK$, denoted $Y^k_i$ as
\begin{equation}
Y^k_i\dist \N(\mu^k,\upsilon^k)
\label{linemodel}
\end{equation}
i.e.,
\begin{equation}
p(Y^k_i|\mu^k,\upsilon^k)=
{1\over\sqrt{2\pi\upsilon^k}} e^{-{(Y^k_i-\mu^k)^2 / 2\upsilon^k}}
\end{equation}
independently for  $i=1,\ldots,N^k$.
Equation~(\ref{linemodel}) represents a line with intensity normalized to one.
The total line counts for a perfect instrument (i.e., with effective
area everywhere equal to its maximum possible value)
are denoted $(N^1,\ldots,N^K)$ and  assumed to be independent
Poisson random variables, 
\begin{equation}
N^k \dist {\rm Poisson} (\tilde\lambda^k) \  
{\rm independently \ for} \  k\in\cK.
\end{equation}

Proper prior information for the lines and the
continuum is important for a reasonable
fit when the spectral model includes emission lines.  
In particular, prior information is especially important for
relatively weak lines, since it is difficult to 
distinguish a weak line from a chance fluctuation in the continuum.
Luckily such prior information is often scientifically forthcoming in
the form of knowledge (e.g., laboratory measurements and physics theory)
of probable sizes and locations of the various
lines. We begin with  the line location and width (actually the variance),
$(\mu^k,\upsilon^k)$ for which priors\footnote{We 
choose this prior distribution partially
because it is the so-called {\it conjugate} prior distribution, i.e., the 
resulting posterior distribution is from
the same family as the prior distribution (e.g., Gaussian with
updated parameters).  This property significantly simplifies
model fitting with no cost in terms of  the accuracy of parameter 
estimation.} 
are assigned independently for each line $k\in\cK$,
\begin{equation}
\upsilon^k \dist {\nu_0^k\upsilon_0^k\over\chi^2_{\nu_0}} 
 \  {\rm and } \ 
p\l(\mu^k|\upsilon^k\r) = \N\l(\mu_0^k,{\upsilon^k\over\kappa_0^k}\r),
\label{eq:lineprior}
\end{equation}
where $\chi^2_{\nu_0}$ is a variable that follows the $\chi^2$ 
distribution with $\nu_0$ degrees of freedom.
We interpret the hyper-parameter, $\bphi^k=(\mu_0^k, 
\upsilon_0^k,\kappa_0^k,\nu_0^k)$ using the mean and variance
of the distributions in Equation~\ref{eq:lineprior}. For example,
\begin{equation}
\E(\upsilon^k|\bphi^k)=\nu_0^k\upsilon_0^k/(\nu_0^k-2) \  {\rm for}
\ \nu_0^2>2
\end{equation}
and
\begin{equation}
\Var(\upsilon^k|\bphi^k)=2(\nu_0^k\upsilon_0^k)^2(\nu_0^k-2)^2(\nu_0^k-4)
\ {\rm for} \ \nu_0^2>4.
\end{equation} 
(Recall that the units here are keV for means and
keV$^2$ for variances.) Thus, the mean and variance of the prior for 
$\upsilon^k$ may be tuned using $\upsilon_0^k$ and $\nu_0^k$;
a small value of $\nu_0^k$ results in a wide, relatively
non-informative prior. 
Since the data are discrete, {\it a priori} we cannot allow the standard 
deviation  of
the line to  become too small (say below the PHA bin width of the bin that
contains the center of $\mu^k$) since there is not information in the
data about the width of a line which is narrower than one PHA channel.
This is accomplished by truncating the prior distribution of $\upsilon^k$.
For the prior on $\mu^k$, the mean and variance are given by 
$\mu_0^k$ and $\kappa_0^k$;  $\mu_0^k$ is the most
probable location of the $k$th line and $\kappa_0^k$ calibrates the
uncertainty in the location of the $k$th line relative to the width of the
line. 

An alternative interpretation of the priors is in terms of additional
hypothetical
photons. Heuristically, the effect of the prior on $\mu^k$ if
$\upsilon^k$ were known would be the same as $\kappa_0^k$ photons
all known to be from line $k$ and equal to $\mu_0^k$. Likewise, the
effect of the prior on 
$\upsilon^k$ is the same as adding $\nu_0^k$ photons with
average squared deviation from the mean equal to $\upsilon_0^k$. 

We now turn to the prior distribution on $\tilde\lambda^k$ and 
set $\tilde{\lambda}^k \dist 
\gamma(\phi^k_1,\phi^{\tilde\lambda}_{2})$,\footnote{We choose a   
$\gamma$  prior partially because it is conjugate to the Poisson 
distribution.}   
(independently)
which has mean $\phi^k_1/\phi^{\tilde\lambda}_2$ and variance
$\phi^k_1/(\phi^{\tilde\lambda}_2)^2$. 
Roughly speaking, the $\gamma$ prior contains the same information as
$\phi_2^{\tilde\lambda}$ Poisson observations (with exposure equal to 
the source exposure) with a total of $\phi_1^k -1$
counts. Since the data consist of a single observation (for each
bin), $\phi_2^{\tilde\lambda}$ can be  interpreted
as the weight put on the prior relative to the data; $\phi^{\tilde\lambda}_2=1$
induces a prior as influential as the data in the absence of absorption,
blurring, background, and lines.  Thus, values of
$\phi^{\tilde\lambda}_2 \ll 1$  are typically recommended for
non informative priors.
The hyperparameters $\phi^k_1$, can be interpreted as the prior 
relative sizes of the lines.  That is, $\phi^k_1 / \sum_{\tilde
  k}\phi^{\tilde k}_1$
is the prior proportion of line photons from line $k$.
We define
the hyper-parameter for line $k$ as $\bphi^k=
(\mu_0^k,\kappa_0^k,\nu^k,(\sigma^2_0)^k,\phi_1^k,\phi_2^{\tilde\lambda})$;
the last element is not indexed by $k$ since it is
constant for $k\in\cK$.

\subsection{Absorption and Correction for Effective Area}

From the viewpoint of our statistical algorithm, 
both the telescope
effective area and astrophysical absorption (e.g., 
absorption due to the ISM) are
handled in the same way. These two processes act independently on 
individual photons and randomly prevent a (energy dependent) proportion of 
photons from being observed. The only essential statistical difference 
being that only the absorption process has unknown parameters.  
In particular, we suppose that the probability a photon is {\it not} 
absorbed (statistically speaking ``censored'') 
by either of these two processes is
\begin{equation}
d_j u(\btheta^A,E_j),
\ {\rm where} \ \log u(\btheta^A,E_j)=\bX_j^A\btheta^A \ 
{\rm for} \  j\in\cJ,
\label{eq:abs}
\end{equation}
where $\btheta^A$ is a $(P^A \times 1)$ parameter, $\bX^A_j$ is a
$(1\times P^A)$ vector  of independent variables, and $P^A$ is the
number of parameters in the absorption model, $u(\btheta^A,E_j)$.
As an example, simple exponential absorption can be written in 
this linear form with 
$\theta^A$ a scalar and $\bX_j^A = -1/\E_j$, i.e.,
$u(\theta^A,E_j)=e^{-\theta^A/E_j}$.  For more complicated 
absorption models, $\bX^A_j$ typically consists of a tabulated absorption function.

The prior for $\btheta^A$ is multivariate Gaussian, $\theta^A_p\dist
\N(\mu_p^A,\upsilon_p^A)$, independently for $p=1,\ldots,P^A$. The prior
is interpreted similarly to that for the continuum parameter $\btheta^C$.
We, however truncate this prior to ensure $\exp\{\bX_j^A\btheta^A\} <1$
for each $j$, to ensure the proportion of photons not absorbed is
less than 1.  With appropriately chosen $\bX_j^A$, this can be
accomplished by assuming each 
component of $\btheta^A$ is negative.

\subsection{Background}

We assume the availability of a separate observation containing background 
counts which can be used to model the background spectrum and correct the
source spectrum. 
Rather than simply subtracting off (a scalar multiple of) the
background counts, however,
we account for the variation due to the Poisson character of the
counts. In particular, we suppose the background count in PHA channel
$l$ is
\begin{equation}
Y^B_l\dist {\rm Poisson}(\theta^B_l)
 \  {\rm independently \ for} \  l\in\cL,
\label{eq:backgd}
\end{equation}
where the unobserved quantity, $Y_l^B$ is the counts in PHA
channel $l$ that are due to background. We parameterize the 
prior for $\theta^B_l$  as $\gamma(\phi^B_{l,1},\phi^B_{l,2})$,
which we expect to be informative based on a pure background
exposure. In particular, a reasonable prior based on 
$Y^{{\rm obs}, B}_l$ background counts would be $\gamma
(Y^{{\rm obs}, B}_l +1,\tau)$,
where $\tau$ is the background exposure time and area relative to the source
exposure time and area. If $Y^{{\rm obs}, B}_l=0$, we generally replace it by
some small number, such as 0.25 in the prior. (This value should not
be too small, since this would preclude the possibility of background counts 
in the corresponding channel in the same exposure.)    

In an extreme case, when the background is very well determined, e.g., via
a very long exposure, we may fix
$\theta^B_l = Y^{{\rm obs}, B}_l/\tau$ and discard the prior
distribution; here we are effectively setting the prior variance to zero.
Note this is not equivalent to subtracting off the background because
we still allow for Poisson variability in the background counts that
contaminate the source counts.
An alternative strategy is to fit a parameterized model to the background. 
For example, we might assume $Y_l^B \dist {\rm Poisson}
(\lambda^B_l(\btheta^B,
\tilde{E}_l))$, where $\log \lambda^B_l(\btheta^B, \tilde{E}_l)=
\bX_l^B\btheta^B$
for $l\in\cL$ with $\bX_l^B$ a row vector of independent variables
depending on the energy of PHA channel $l,\tilde{E}_l$.  This allows the 
background counts to be modeled as
a power law, broken power law, or any other log linear model.

\section{Applications}

In this section, we illustrate our methods and algorithms using two
datasets. We first analyze
{\it ASCA}/SIS data of high redshift (z=3.384) quasar S5~0014+813
\citep{elvi:etal:94} to illustrate the various summaries available in
a relatively straight-forward MCMC analysis. 
The second analysis involves an extremely low-count stellar coronal
source ($\alpha$ TrA) and illustrates the power of Bayesian methods to
combine information from various sources and quantify the weak
information available in this data.

\subsection{Quasar S5~0014+813} 

A typical quasar X-ray spectrum can be described by an absorbed power
law. A fluorescent iron line (Fe K-$\alpha$) emitted at energy between
$\sim 6.4-6.8$~keV if detected can be a signature of a reflection
component and its ionization state \citep{geor:etal:00}.
Quasar S5~0014+813 \citep{kuhr:etal:81} at redshift z=3.384 is among
the highest X-ray flux quasars known with z $\sim$ 3.
S5~0014+813 was observed with {\it ASCA} on 1993 October 29 with an
exposure time of 22.8 ks in the SIS0 detector \citep{elvi:etal:94}. 
Here we apply our model to this data to illustrate the method and look
for signatures of the iron emission line.

The spectral data were extracted with the standard screening criteria
\citep{elvi:etal:94} and standard response matrices were used
(ftp://legacy.gsfc.nasa.gov/caldb/data/asca/sis/cpf/94nov9).  We use
all of the original 512 PHA instrument channels except the unreliable
channels below $\sim 0.5$ keV and above $\sim 10$keV.  In addition we
do not group any channels. (Channels are usually grouped in order to
justify the use the default $\chi ^2$ techniques with 
their Gaussian assumptions.) As is allowed with a Poisson model, we
instead  use only the original PHA bins.
The source model included the exponential shape of Galactic absorption
(see Section~3.5), and a power law continuum (i.e.,
Equation~\ref{eq:powerlaw}) with a narrow emission line at 1.45~keV
(observed frame; $\sim 6.7$keV rest frame).  We accounted for
background using a background Poisson process with intensity equal to
the (rescaled) background counts in each PHA channel. Flat priors were
used on all model parameters.

To estimate the four model parameters (i.e., the power law, 
normalization, and exponential absorption parameters and the equivalent 
width\footnote{The equivalent width is defined as
$\tilde\lambda^k/f(\th^C,\mu^k)$.}
of the line) a sample from their posterior distribution was 
obtained by running three MCMC chains using dispersed starting values. 
The chains showed excellent mixing (as measured with $\sqrt{\hat R}$, see
Appendix~A.2) after 2000 draws. In the Monte Carlo evaluation, the
second half of each of the chains were used along with an additional run
of 2000 draws from each chain, for a total of 9000 draws.

Summaries of the model fit appear in Table~\ref{tbl:quasar},
Figure~\ref{fig:quas-bipost}, and  Figure~\ref{fig:quas-fit}.
The parameter estimates are posterior means computed using a
transformation which makes the marginal posterior 
distributions more symmetric and hence the posterior mean a more
informative summary  (i.e., ln(Normalization) and
sqrt(Equivalent Width)).
In particular, if we represent the draws of
the normalization parameter as $\{\th_{[t]},t=1,\ldots,9000\}$,
the point estimate of this parameter was computed as
\begin{equation}
\exp\l\{{1\over 9000}\sum_{t=1}^{9000}\ln(\th_{[t]})\r\} =
\root 9000 \of{\prod_{t=1}^{9000} \th_{[t]}},
\end{equation}
the geometric mean. The credible intervals are computed using the 0.025
and
0.975 quantiles of the draws and are invariant to (monotonic)
transformations. Pairwise credible regions appear in
Figure~\ref{fig:quas-bipost}. The scatter plots illustrate the regions
of highest posterior probability by plotting the Monte Carlo draws ---
$\Pr(\btheta\in R)$ is
approximately equal to the proportion of points in that region.
The grey-scale images give Monte Carlo estimates of the (darker) 50\% and
(lighter) 90\%  marginal posterior regions. The grainy character of the
images is due to the Monte Carlo approximation. Even with this
relatively large data set and with the use of 
transformations, the non-Gaussian character of the posterior is evident.
We expect that higher dimensional marginal posterior distributions 
are even less
Gaussian in character. Figure~\ref{fig:quas-fit} compares the fitted
source model corrected for effective area and absorption with the PHA
counts and illustrates the estimated continuum and the stability of
this estimate.





\subsection{Hybrid-Chromosphere Supergiant Star $\alpha$ TrA}

Unlike the simple power-law spectrum of the quasar in the previous
section, stellar coronal spectra are complicated by a Bremsstrahlung
continuum and the presence of numerous emission lines.  Such complex
spectra are much more difficult to model, and in addition, the
intensity of the Bremsstrahlung continuum drops exponentially at
high-energies, resulting in very few counts.  Analyzing such spectra
is however crucial to the understanding of coronal structure, mechanisms
of coronal heating, etc.  A case in point is the corona of the
Hybrid supergiant star $\alpha$ TrA (HD\,150798, K4II, B-V=1.44,
V=1$^m$.92), which shows evidence of both strong magnetic activity
as indicated by X-ray emission \citep{brow:etal:91, kash:etal:94}
and stellar outflow seen in absorption profiles \citep{hart:etal:81}.
X-ray observations with the {\sl ROSAT/PSPC} \citep{kash:etal:94}
indicate that its corona is dominated by transient, unstable plasma
that is confined by magnetic loops that are closed on short length
scales \citep{rosn:etal:94}.  Constraining the maximum temperatures present
in the corona is therefore of primary importance.  Here we use data
obtained with {\sl ASCA}, at higher energies than {\sl ROSAT}, to model
the spectrum.  The low number of counts detected at high energies make
this spectrum difficult to analyze by traditional means, and we must
bring to bear the full power of a hierarchical Bayesian analysis in
order to constrain the maximum temperatures present in the corona.



$\alpha$ TrA was observed with {\sl ASCA} in March 1995 for
$\approx$~34~ks.  During this observation, the source exhibited no
flares. The count rate was steady, and corresponded to the
quiescent state identified with {\sl ROSAT}.

We model the high-energy region of the {\sl ASCA} spectrum (2.5 -- 7.5 keV)
as a combination of a Bremsstrahlung continuum
\begin{equation}
\frac{Norm}{\sqrt{T}} e^{-E/k_B T} \,,
\end{equation}
where $T$ is the electron temperature, $E$ is the energy in keV, and
$k_B$ is the Boltzmann constant, and
$Norm$ is a normalization; and a number ($\sim 10$) of narrow emission lines
located at the positions of known strong lines whose widths and
locations\footnote{Line
location can be known only to the resolution of the instrument, and
hence each of the model lines represents the sum of a large number of
lines within the resolution element; we find that we only exclude
$< 5\%$ of the line flux by this approximation in the energy range
considered.}
are fixed, but intensities are allowed to vary. 
As is our convention, the units of $Norm$ is counts per
keV per maximum effective area.  Because of the low
counts we fit a power law to the background.

We apply the above model to {\sl SIS0} data ($\sim 28$ ks) and the
combined data from the 2 {\sl GIS} detectors ($\sim 33$ ks in each).
Note that the very low counts present in these data ($\sim 150$ counts
in {\sl SIS0}, $\sim 300$ in {\sl GIS}) preclude any ``traditional''
analysis: it is only by using the full Bayesian machinery that we can
derive useful results from such data.

We carry out the analysis in two steps:
\begin{enumerate}
\item Choose highly non-informative priors 
on the parameters to analyze {\sl SIS0} data: for the prior on normalization,
$p\left(\frac{Norm}{\sqrt{T}}\right)$, such that it lies in between
$10^{-9}$ and $10^{-1}$ with a 90\% probability;
on temperature, $p\left(\frac{1}{k T}\right)$ such that it is always
positive and also is nearly flat in the temperature range of interest;
and on line intensity, $p\left({\tilde\lambda^k}\right)$ such that
a priori all the lines have the same intensity, and the maximum total
counts due to lines is 100 (the total source+background counts for the
SIS0 observation is only 154; atomic emission line models indicate that
for the temperature and energy range of interest, 100 corresponds to the
maximum possible contribution to the spectrum from lines).
We choose Gaussian forms for the first two and a Gamma prior for the last;
these priors are illustrated as solid lines in Figure~\ref{fig:alpha-s0-post}.
Thus,
\begin{eqnarray}
p\left(\ln\left(\frac{Norm}{\sqrt{T}}\right)\right) & = & 
N(\mu=-9.21,\sigma=5.58) \label{eq:vinayPrA}\\
p\left(\frac{1}{k T}\right) & = &  N(\mu=0.95,\sigma=2.5) \label{eq:vinayPrB}\\
p\left({\tilde\lambda^k}\right) & = & 
\gamma(\phi_1^k=0.11,\phi_2^{\tilde\lambda}=0.0033) \ {\rm for} \
k=1,\ldots 10 \,. \label{eq:vinayPrC}
\end{eqnarray}
\item Use the posterior resulting from the above step to define 
more informative priors to analyze {\sl GIS} data.
These priors also correct for the difference in
exposure time and average effective area between the SIS0 and the GIS data.
The posterior variances from the initial analysis were increased
somewhat when computing the priors for the second analysis. These
priors are illustrated as solid lines in Figure~\ref{fig:alpha-gis-post}.
Thus,
\begin{eqnarray}
p\left(\ln\left(\frac{Norm}{\sqrt{T}}\right)\right) & = & 
N(\mu=-9.97,\sigma=1.74)\\
p\left(\frac{1}{k T}\right) & = & N(\mu=0.41,\sigma=0.43)\\
p\left({\tilde\lambda^k}\right) & = & 
\gamma(\phi_1^k=0.12,\phi_2^{\tilde\lambda}=0.025) \ {\rm for} \
k=1,\ldots, K \,.
\end{eqnarray}
\end{enumerate}
We ran three Markov Chains in each analysis to obtain draws from the 
posterior distribution of $\l(\ln(Norm/\sqrt{T}), 1/kT,
\tilde\lambda^1,\ldots, \tilde\lambda^{10}\r)$. In both analyses there was
excellent mixing after 6000 draws and we used the second half of each
chain for a total of 9000 Monte Carlo draws.

The results of the analysis are shown in
Figures~\ref{fig:alpha-s0-post} -- \ref{fig:alpha-fit}.
(In the figures the parameter $\Omega$ refers to the proportion of
source photons from the lines and 
$\lambda=\sum_{k=1}^{10} \tilde\lambda^k$.)
We find that the plasma temperature is $\sim 19_{>11}^{<64} \times 10^6$ K
(Figure 5).  Such a large value (cf.\ $\sim 2 \times 10^6$ K in the quiet
Solar corona) clearly lends credence to the idea that the corona on
$\alpha$ TrA is dominated even in quiescence by flare-like events.

As a byproduct of our analysis, we also obtain the flux in the modeled
lines relative to the continuum.  In principle, this allows us to
constrain the metallicity {\it for the first time} in the corona
of $\alpha$ TrA by comparing the observed ratio of the
line and continuum fluxes\footnote{Incompleteness in atomic line databases
(\cite[see][]{bric:etal:98}) contribute to an error of $<5\%$ on
the line-to-continuum ratios calculated here.  They are negligible
compared to the measurement error.}
with that derived from thermal emission models computed over the
same temperature range \citep{drak:etal:98, kash:etal:98}.
The photospheric metallicity \citep{tayl:99} is [Fe/H] = 0.3, and
we derive for the coronal metallicity [Fe/H] = $0.4_{-0.6}^{+1.1}$
where the quoted range represents posterior deviations at $1~\sigma$.  While 
the
uncertainty on our measurement is quite large (it is essentially unbounded
at high metallicity), it is encouraging that the corona does not appear
to be metal abundance deficient \citep[see][]{drak:96}.

\section{Discussion}
The power of the Bayesian methods illustrated here lies in their
ability to combine information and to directly model the highly
structured hierarchical features of the data---both in a principled
manner. These features are illustrated in the $\alpha$ TrA
example. First, by combining information from several detectors,
we are able to extract information
from the data regarding the plasma temperature. More generally,
Bayesian methods allow for the incorporation of various forms
quantifiable prior information through the prior distribution. Of
course, results are then conditional on the prior 
information---if these priors are not trusted, the conclusions 
cannot be trusted
either. On the other hand, if the prior information is accepted as
reasonable, the posterior distribution should be
accepted as a conglomeration of prior scientific information and the
data. Second,
the extremely low counts in the $\alpha$ TrA  data, along with many free
parameters (ten emission line intensities and two continuum
parameters) illustrate a situation in which methods based on the
Gaussian distribution and the central limit theorem are simply without
justification. Methods which account for the Poisson (e.g., highly
variable) character of the data have a sound mathematical basis and, 
in contract to standard methods such as $\chi^2$ fitting, are
equipped to handle such data.  

The hierarchy in the model described in Section~3 can be extended to
account for various more complicated features in the data, e.g.,
absorption lines, pile-up, and joint spatial, spectral, and temporal
structure. Dealing with pile-up is perhaps the most important outstanding
data-analytic challenge for Chandra. Conceptually, however, there is
no difficulty  in addressing pile-up in a Bayesian framework. After
accounting for other features in the data such as instrument response,
background, and absorption, we simply need to separate the observed
counts into multiple counts of lower or equal energy based on the
(current draw of the) spectral and spatial model. The difficulty lies
in computation. Simply enumerating the set of photons that could
result in a particular observed event, let alone their relative
probabilities is an enormous task. Thus, we believe there is
great promise in Monte Carlo techniques which if carefully designed,
can automatically exclude numerous possibilities with minute
probability. Although there remains much work to be done, Bayesian
methods in conjunction with MCMC algorithms offer a practical and
innovative solution to many outstanding data-analytic challenges in
astrophysics. 

\if1\aastex
\acknowledgements  
   The authors gratefully acknowledge
    funding for this project partially provided by NSF grant
    DMS-97-05157, the Chandra X-ray
    Observatory, and by NASA grant NAS8-39073. They also thank Rostislav
    Protassov  and
    C.J. Shen for their work on the programming of the various algorithms,
    Richard Edgar and Paul Gorenstein for helpful comments on an earlier 
    draft, and the referee, Jeff Scargle, whose careful reading and detailed
    suggestions have greatly improved the paper.
\fi

\clearpage

\appendix
\noindent{\bf \huge{Appendices}}

\section{Markov Chain Monte Carlo Methods} 

\subsection{The Data Augmentation Algorithm}

The Data Augmentation algorithm is designed to obtain a sample
from the posterior distribution for use in Monte Carlo integration.  
The strategy of the algorithm is to embed the posterior distribution,
$p(\btheta|\bY)$, into a distribution in a large space,
$p(\btheta,\bY\mis|\bY)$.  If we can obtain a sample from
this second distribution, we need only discard the sampled
values of$ \bY\mis$ to obtain the desired sample from the
posterior. The quantity $\bY\mis$ can be any unobserved
quantity; it is referred to as ``missing data'' for historical 
reasons.  For clarity we denote the
observed data $\bY\obs$ and the augmented data
$\bY\aug=(\bY\obs,\bY\mis)$.  In order to obtain a
sample from $p(\btheta,\bY\mis|\bY\obs)$, the
Data Augmentation algorithm uses an iterative sampling
scheme that samples first $\bY\mis$ conditional on the model
parameters and $\bY\obs$ and second samples the model parameters 
given $\bY\aug$.  Clearly, the algorithm is 
most useful when both of these conditional
distributions are easily sampled from. The iterative character of the
resulting chain naturally leads to a Markov chain, which we initialize at
some starting value, $\btheta_{[0]}$. For $t=1,\ldots,T$,
where $T$ is dynamically chosen, we repeat the following two steps: 

\smallskip
\begin{description}
\item{Step 1:} Draw $\bY\aug_{[t]}$ from $p(\bY\aug|\bY\obs, \btheta_{[t
    -1]})$, 

\item{Step 2:} Draw $\btheta_{[t]}$ from $p(\btheta|\bY\aug_{[t]})$. 
\end{description}
\smallskip
\noindent
Under certain regularity conditions \citep[see][for details]{meyn:twee:93,
robe:96,tier:94,tier:96} 
the stationary distribution of the resulting Markov chain is
the desired posterior distribution, i.e., for large $t$, $\btheta\cur$
approximately follows $p(\btheta|\bY\obs)$.

To illustrate the utility of the Data Augmentation algorithm, 
we return to the simple background contamination
model introduced in Section~2.1.  The choice of $\bY\aug$ 
is clear in the example; we set $\bY\aug=\{Y,Y^S,
Y^B\}$, where $Y$ is the total counts, $Y^S$ is
the unobserved source counts from the source exposure, and $Y^B$ is
the counts from the pure background observation. (I.e., we can
consider $Y^S$ to be the missing data.) With
this choice of $\bY\aug$, both $p(\bY\aug|\bY\obs, \btheta)$ and 
$p(\btheta|\bY\aug)$
are easy to sample and thus the Data Augmentation algorithm is easy to
use; here $\bY\obs=\{Y,Y^B\}$ and $\btheta=(\lambda^B,\lambda^S)$.
Given some $\btheta_{[0]}=(\lambda^B_{[0]}, 
\lambda^S_{[0]})$ the two steps of
the Data Augmentation algorithm at iteration $t$ become

\smallskip
\noindent
{Step 1:} Draw $Y^S\cur$ from 
\begin{equation}
Y^S\Big|\btheta\pre,\bY\obs\ \dist \
{\rm binomial}\l(Y,{\lambda^S\over\lambda^B+\lambda^S}\r),
\label{eq:exDRAWa}
\end{equation}
i.e.,
\begin{equation}
p\l(Y^S\Big|\btheta\pre,\bY\obs\r) = {Y\choose Y^S}
(Y^S)^{\lambda^S/(\lambda^B+\lambda^S)}
(Y-Y^S)^{\lambda^B/(\lambda^B+\lambda^S)}
\end{equation}

\noindent
{Step 2:} Draw
\begin{equation}
\lambda^B\cur\Big|\bY\aug\cur \ \dist \
\gamma\l(\alpha^B+Y-Y^S, \beta^B+1\r),
\label{eq:exDRAWb}
\end{equation}
i.e., 
\begin{equation}
p\l(\lambda^B\cur\Big|\bY\aug\cur\r)=
\l(\lambda^B\cur(\beta^B+1)\r)^{(\alpha^B+Y-Y^S)}
e^{-\lambda^B\cur(\beta^B+1)}
\big/(\lambda^B\cur\Gamma(\alpha^B+Y-Y^S)),
\end{equation}
where $\alpha^B$ and $\beta^B$ are typically chosen using the pure
background observation as described in Section~2.1, and
\begin{equation}
\lambda^S\cur\Big|\bY\aug\cur \ \dist \
\gamma\l(\alpha^S+Y^S, \beta^S+1\r).
\label{eq:exDRAWc}
\end{equation}
\smallskip
\noindent
In the first step, we stochastically divide the source
count into source counts and background counts based on 
the current values of $\lambda^B$ and $\lambda^S$.  In
the second step we use this division to update $\lambda^B$
and $\lambda^S$.  Markov chain theory tells us the
iteration converges to the desired draws from the
posterior distribution.

By selecting a starting value and iteratively sampling 
according to Equations~\ref{eq:exDRAWa}, \ref{eq:exDRAWb}, 
and \ref{eq:exDRAWc}, we obtain a Markov chain which delivers
a dependent sample from the posterior distribution upon convergence.
In the next section, we use the Data Augmentation algorithm to 
illustrate the important practical issues of selecting starting values, 
detecting convergence, and accounting for the dependency in the
sample.

\subsection{Starting Values, Convergence, and Multiple Chains}

An important and difficult aspect of MCMC  methods in practice is
ascertaining convergence to stationarity. Since the stationary 
distribution of the Markov chain is the posterior distribution 
on interest, we can consider $\{\btheta\cur, t > T_0\}$
to be a (dependent) posterior sample, which can be used for Monte
Carlo integration. Thus, determining $\btheta_{[0]}$ and $T_0$ is critical
for valid inference. There is a large and growing literature on these
related subjects and we refer interested readers 
to recent texts on the subject
by \citet{gelm:carl:ster:rubi:95}, \citet{carl:loui:96}, and
\citet{gilk:rich:spie:96}, as well as the review article on convergence
by \citet{cowl:carl:96}. Here we briefly outline the approach that we
find most fruitful. 

As proposed by \citet{gelm:rubi:92}, we suggest
running multiple Markov chains with a variety of
starting values spread throughout the parameter space. This is a
useful procedure since a single chain can appear to have converged
when actually it has only settled temporarily in one region of the
parameter space. This is illustrated with the Markov chain in
Figure~\ref{fig:slowconvg}. The three chains show the draws of
a variance parameter for a random-effects model 
\citep[for details see][]{vand:meng:00art}. Note 
that although chain 3 appears relatively stable it
is far from convergence during the first 10,000 draws. This is evident
when it is compared with the other chains, but less so when we look
only at the beginning of 
chain~3. It is recommended that the starting values 
for the several chains be spread
broadly   in the parameter space (relative to the region of high
posterior probability). This can often be accomplished by roughly
mapping the posterior, for example,
using estimates and errors based on the chi square
estimates, posterior modes, or maximum likelihood estimates. (See
\citet{vand:00astro} for details on the computation of posterior
modes and maximum likelihood estimates for our spectral model.) Once such
``over dispersed'' starting values are obtained, we can run the
several chains until all converge to the same region of the
parameter space. (There may be more than one mode in the posterior, in
which case the chains may converge to different modes,
i.e., different regions of the parameter space.)
\citeauthor{gelm:rubi:92}'s (\citeyear{gelm:rubi:92})
$\sqrt{\hat R}$ statistic measures the relative size
of the total variance in the draws of a univariate function of the
parameter and the average within chain variance of the same function, i.e.,
\begin{equation}
\sqrt{\hat R} = \sqrt{\frac{\frac{T-1}{T}W + \frac{1}{T}B}{W}},
\end{equation}
where $B$ is the between chain variance, 
$W$ is the within chain variance, and $T$ is the number of draws.
If the variance within each chain is as great as the total variance
in all the draws, i.e., $\sqrt{\hat R}$ is near one, then we can be
confident that all the chains have converged to the same region of the
parameter space.  Typically we compute  $\sqrt{\hat R}$ using the
last half (or two thirds) of each of the chains. Once an acceptable level
of  $\sqrt{\hat R}$ is obtained (say below 1.2) we omit the first half
(or third) of the chain in all further analysis.
If we have several starting values which cover a large
enough region of the parameter space, we can be confident that the
chains sample all areas with high posterior probability and thus
the Monte Carlo approximations are unbiased estimators of the
quantities they estimate.

The variance of the Monte Carlo
approximations is a function of the posterior variance of the
quantity being approximated, the posterior sample size (i.e., $T-T_0$), and 
the autocorrelation function of the Markov chain. Typically Monte
Carlo errors are small relative to the posterior variance
with several thousand posterior draws and thus are of little
consequence.  Monte Carlo error can be quantified 
by repeating the analysis for the first half and second half
of the Markov chain and noting if the results are substantively different.
See \citet{robe:96} and the
references therein for details and extensions.

\subsection{The Gibbs Sampler}

In this and the next section we describe two additional MCMC
methods which are designed to delivers a sample from the posterior 
distribution and are often useful when the Data Augmentation 
algorithm is not practical.  The Gibbs sampler 
can be viewed as an extension of the Data Augmentation algorithm
in which we wish to sample from $p(\btheta|\bY\obs)$, and the vector,
$\btheta$, can be viewed as a combination of model
parameters and ``missing data''.  (In many instances, there
is no ``missing data''.)  We partition $\btheta$ into
$(\btheta_1,\ldots,\btheta_P)$, where $\btheta_p$ may be a scalar
or vector quantity for each $p$.  The Gibbs sampler, again starts with
some starting value $\btheta_{[0]}$ and at iteration $t$ samples
according to the following conditional distributions:

\smallskip
\begin{description}
\item{Step 1:} Draw 
$(\btheta_1)\cur\dist p(\btheta_1|(\btheta_{-1})\cur, \bY\obs)$,
\item{Step 2:} Draw 
$(\btheta_2)\cur\dist p(\btheta_2|(\btheta_{-2})\cur, \bY\obs)$,
\item{$\quad\quad\vdots\quad$}
\item{Step $P$:} Draw 
$(\btheta_P)\cur\dist p(\btheta_P|(\btheta_{-P})\cur, \bY\obs)$,
\end{description}
\smallskip
\noindent
where  $(\btheta_{-p})\cur=
((\btheta_1)\cur,\ldots,(\btheta_{p-1})\cur,(\btheta_{p+1})\pre,\ldots,
(\btheta_P)\pre)$.
That is, we draw each component of $\btheta$ in turn conditional on
the current values of the rest of $\btheta$ and the data. 

The advantage of the Gibbs sampler over the Data Augmentation
algorithm is that in many settings additional conditioning results in
simpler draws. The disadvantage is that the resulting Markov chains
tend to have higher autocorrelation and are slower to converge
to stationarity as $P$, the number of steps per iteration, increases.

\subsection{The Metropolis-Hastings Algorithm}

As a final extension, we consider the case when one (or more) of the
steps in the Gibbs sample involves a conditional distribution that is
not easy to sample. The Metropolis-Hastings algorithm \citep{metr:ulam:49,
metr:rose:rose:tell:tell:53, hast:70} replaces the
conditional distribution by some convenient ``jumping rule'' which
approximates the conditional distribution.  A proposal draw is
sampled according to the jumping rule and is either 
accepted or rejected (in which
case, the Markov chain is fixed at the previous draw) according to a rule that
maintains the desired stationary distribution 
\citep[see, e.g.,][for details]{gelm:carl:ster:rubi:95}.

\section{Details of the MCMC Algorithm}

\subsection{Data Augmentation}

The algorithms used to fit the model 
described in Section~3.1 rely on the method of data
augmentation.  In this section, we detail the layers of the 
data-augmentation scheme we use.  We aim to construct an idealized data 
set for which model fitting is a relatively easy task.  
That is, given the augmented data, we can easily sample the model 
parameters. Likewise, given the model parameters, we can easily sample 
the augmented data and thus we can construct a Data Augmentation
algorithm as described in Appendix~A.1.
Suppose, for example, 
a data set uncontaminated by background or instrument 
response were available. Clearly, model fitting would be easier.  We
define an even  
larger data set that contains the unbinned, true, and blurred energies of 
all photons that would have arrived at the detector if there had 
been no absorption and if we were using a perfect instrument
with the effective area equal to its maximum
value over all energies used.  This data set also includes a variable 
indicating absorption and loss to reduced effective area\footnote{
Absorption and effective area are handled together, so we need
only one indicator variable, see Section~3.4.}
and a variable 
indicating the source of each photon, i.e., background ($B$), 
continuum ($C$), and each of the $K$ line profiles; the set of sources is 
denoted $\cS=\{B,C,1,2,\ldots,K\}$.  This idealized data set is
summarized in Table~\ref{tbl:data}.

The data augmentation scheme
is illustrated in Figure~\ref{fig:model} in which squares and circles 
represent observed and unobserved (``augmented'') quantities, 
respectively. Given the model parameter $\btheta$, we obtain a sample 
set of photon
energies, $\Yc^s=(\mYc^s_1,\ldots,\mYc^s_{N^s})'$ for $s\in
\cS$ (see the third column of Figure~\ref{fig:model}) represented the
undegraded ``augmented'' data; $N^s$ is the total count for source $s$. 
(As a mnemonic device, more dots in the accent 
above $Y$ signifies further removal of a quantity from
actual observable quantities.)
Here, $\Yc^k$ contains the exact energy of all photons attributed 
to line $k$ before absorption, with maximum effective area, and no 
background contamination. 
(The background photon energies, $\Yc^B$, do not appear in 
Figure~\ref{fig:model} because 
we model the detected counts (e.g., in PHA channels) rather 
than true counts, see Section~3.6.)  The first two columns of
Figure~\ref{fig:model} represent the hyperparameters and model parameters
detailed in Sections~3.2--3.5.

The array of energies represented by $\Yc^s$ are binned into instrument-specific
energy bins to obtain a sample spectrum, $\Yb^s=(\mYb_1^s,\ldots,\mYb_J^s)'$
(see the fourth column in Figure~\ref{fig:model}). In particular,
\begin{equation}
\mYb^s_j=\sum\itoNs I\{\mYc^s_i\in B_j\}, \  
{\rm for}  \  j\in\cJ \   {\rm and}  \   s\in \cS,
\end{equation}
where $B_j$ is the $j$th energy bin. The first plot in
Figure~\ref{fig:datagen} illustrates the undegraded counts from 
the continuum and lines,
$\mYb\updot_j=\sum_{s\in\cS\setminus B}\mYb^s_j$ for the artificial
data set, where the notation $\cS\setminus B$ indicates set subtraction, 
i.e., the set $\cS$ with $B$ removed.
The solid line represents $\E(\mYb_j\updot|\btheta)$ which
equals the term in square brackets in Equation \ref{eq:model_true}.
Due to absorption  
and effective area, a portion of these
photons are not detected. The sample counts {\it after} absorption 
(and accounting for effective area) are
depicted in the fifth column of Figure~\ref{fig:model}, by
$\Ya^s=(\mYa_1^s,\ldots,\mYa_J^s)'$ with
\begin{equation}
\mYa_j^s=\sum\itoNs I\{\mYc_i\in B_j\}(1-Z_i^A)
 \ {\rm for}  \  j\in\cJ  \  {\rm and}  \  s\in\cS;
\end{equation}
as described in
Table~\ref{tbl:data_summaries}, $Z^A_i$ is one if photon $i$
is absorbed and is zero otherwise.
The second plot in Figure~\ref{fig:datagen} represents $\Ya^s$ with
$\E(\mYa_j\updot|\btheta)=\lambda_j(\btheta)d_j$
plotted as the solid line; see Equation~\ref{eq:model_true}.
The next two circles in
Figure~\ref{fig:model}  represent the
adding of sources and the blurring (i.e., instrument response) process. In
particular, $\Ya\updot=\sum_{s\in\cS\setminus B}\Ya^s$. The blurred 
data, 
$\bY\updot=(Y\updot_1,\ldots,Y\updot_L)'$, is a stochastic function  of
$\Ya\updot$, 
(i.e., a multinomial distribution\footnote{The
multinomial $(n,\bp)$ distribution is a distribution for nonnegative
integer valued random vectors and generalizes the binomial
distribution.  In particular, a vector randomly selected from this 
distribution sums to $n$ and
its expected value is $np$, where $p$ is a probability vector
which sums to one.})
\begin{equation}
\bY\updot\dist\sum_{j\in\cJ}{\rm multinomial}(\mYa_j\updot,\bM_j),
\end{equation}
where $\Ya\updot=(\mYa_1\updot,\ldots,\mYa_J\updot)'$ and $\bM_j$ is the
$j$th column of $\bM$; $\bY\updot$ appears in the third plot of
Figure~\ref{fig:datagen} with 
$\E(Y\updot_l|\btheta)=\sum_{j\in\cJ}M_{lj}\lambda_j(\btheta)d_j$.
 The counts due to background  
contamination are denoted $\Ya^B=(\mYa_1^B,\ldots,\mYa_L^B)'$  
and the
observed data are denoted $\bY\obs=(Y\obs_1,\ldots,Y_L\obs)'$ with
$Y_l\obs=Y_l^B + Y\updot_l$  for $l\in\cL$; $\bY\obs$ is illustrated in
the final plot of Figure~\ref{fig:datagen} and has expectation $\xi(\btheta)$;
(cf. Equation~\ref{eq:model_obs}).

\subsection{The Algorithms}

In this section we present the details of the MCMC algorithm
which we use to sample from the posterior distribution for
our spectral model. We  use an algorithm which alternately draws the
``missing data'' given the model parameters and the parameters
given the ``missing data''.  Both draws are conditional on the
observed photon counts and the prior hyperparameters,
$\bphi=\{\bphi^A,\bphi^s,s\in\cS\}$.
In particular, we define two groups,
(1) the augmented data, $\bY\aug=\{\bY\obs,\bY^B,\bY\updot,
\Ya, \Yb, \Yc\}$, where
$\Ya=\{\Ya^s,s\in\cS\setminus B\}$ are the binned true energies, after
absorption and accounting for effective area, $\Yb=\{\Yb^k, k\in\cK\}$
are the binned true energies
and $\Yc=\{\Yc^k,k\in\cK\}$ are the (unbinned) true energies and
(2) $\btheta=\{\btheta^A, \btheta^s, s\in\cS\}$ consists of the various model parameters.
Using Bayes's Theorem, we are able to
derive the necessary conditional distributions, which are
described below. 

First, we draw $\bY\aug$ from $p(\bY\aug|\bY\obs,\btheta)$;
the draw is broken into the following five steps. 

\smallskip
\noindent
{\it Draw 1, Step 1:} Independently separate the background counts,
\begin{equation}
Y^B_l\Big|\bY\obs,\btheta \ \dist \ {\rm binomial}\left(Y\obs_l,
{\theta^B_l\over\xi_l(\btheta)}\right), \ {\rm for} \ l\in\cL.
\end{equation}

\smallskip
\noindent
{\it Draw 1, Step 2:} Restore the blurred the photons,
\begin{equation}
\Ya\updot\Big|  \bY^B, \bY\obs, \btheta \ \dist \
\sum_{l\in\cL} \ 
{\rm multinomial}\l(Y\updot_l,{{d_1\lambda_1(\btheta)M_{1l},
\ldots,d_j\lambda_j(\btheta)M_{Jl})} \over
{\sum_{j\in\cJ}d_j\lambda_j(\btheta)M_{jl}}}\r),
\end{equation}
where $Y\updot_l=Y\obs_l-Y^B_l$.

\smallskip
\noindent
{\it Draw 1, Step 3:} Independently separate the counts into line and
continuum counts, 
\begin{equation}
\l(\Ya^C,\Ya^1,\ldots,\Ya^K\r)\Big| \Ya\updot, \bY^B, \bY\obs, \btheta
\ \dist \ \phantom{put a big big big space here}
\end{equation}
\begin{equation}
\phantom{put a big big big space here} {\rm multinomial}\l(\mYa\updot_j,
{\l(d_jf(\btheta^C,E_j),\tilde{\lambda}^1 p_j^1
\ldots,\tilde{\lambda}^K p_j^K\r)\over
d_jf(\btheta^C,E_j) +
\sum_{k=1}^K \tilde{\lambda}^k p_j^k}\r) \  {\rm for} \ j\in\cJ,
\end{equation}
where $p^k_j=P_j(\mu^k,\upsilon^k)$.

\smallskip
\noindent
{\it Draw 1, Step 4:} Independently restore the absorbed counts in
the lines, 
\begin{equation}
\mYb^k_j\Big|\Ya,\bY^B,\bY\obs,\btheta \ \dist \
\mYa^k_j + {\rm Poisson}\l(\tilde{\lambda}^k
p_j^k(1-d_ju(\btheta^A,E_j))\r),
\end{equation}
for $j\in\cJ$ and $k\in\cK$.

\smallskip
\noindent
{\it Draw 1, Step 5:} Independently de-round the photon energies from the
lines,
\begin{equation}
\mYc^k_{i} \Big|\Yb,\Ya,\bY^B,\bY\obs,\btheta \
\dist \
N(\mu^k,\upsilon^k),  \  \hbox{truncated to} \ B_j
\end{equation} 
for $i=1,\ldots, \mYb_+^k$ and $k\in\cK$ with $\Yb_+^k = 
\sum_{j\in\cJ} \Yb_j^k$.  We note that
this draw is omitted for line $k$ if $\upsilon^k$ is fixed
at zero, i.e., if line $k$ is a delta function.  (In this case
$\mu^k$ is not fit by the algorithm).

Second, we draw $\btheta$ from $p(\btheta|\bY\aug,\bphi)$, taking 
advantage of conditional independence among several
vector components of $\btheta$.

\smallskip
\noindent
{\it Draw 2, Step 1:} Independently draw the background model parameters,
\begin{equation}
\theta^B_l \Big| \bY\aug, \bphi \\ \dist \\
\gamma(\phi^B_{l,1}+\bY^B_l,\phi^B_{l,2}+1).
\end{equation}

\smallskip
\noindent
{\it Draw 2, Step 2:} Draw the variance and mean independently for each
line profile
\begin{equation}
\upsilon^k \Big| \bY\aug, \bphi \\ \dist \\
\phantom{add space add space add space add space add space add space}
\end{equation}

\begin{equation}
\phantom{add space}
{1\over \chi^2_{\nu_0^k +\mYb^k\downdot}}\l[
\nu^k_0(\sigma^2_0)^k
+\sum_{i=1}^{\mYb^k\downdot}(\mYc_i^k-\mYc^k\downdot/\mYb^k\downdot)^2
+{\kappa^k_0\mYb^k\downdot\over \kappa_0^k+\mYb^k\downdot}
 (\mu^k_0-\mYc^k\downdot/\mYb^k\downdot)^2
\r],
\end{equation} 

\begin{equation}
\mu^k \Big| \upsilon^k, \bY\aug, \bphi \\ \dist \\
N\l( {\kappa_0^k\mu^k_0 + \mYc^l\downdot\over \kappa^k_0+\mYb^k\downdot},
{\upsilon^k\over\kappa^k_0+\mYb^k\downdot}\r),
\end{equation}
where $\mYc^k\downdot=\sum_{i=1}^{\mYb^k\downdot}\mYc^k_i$.  Again,
this step is omitted for line $k$ if it is assumed to be a delta
function.

\smallskip
\noindent
{\it Draw 2, Step 3:} Draw the line intensities $\tilde{\lambda}^k$,
independently for each $k$
\begin{equation}
\tilde{\lambda}^k\big| \bY\aug, \bphi \\ \dist \\
\gamma(\mYb^k\downdot + \phi^k_1,
1+\phi^k_{\tilde{\lambda}}) \ {\rm independently \ for} \ k\in\cK.
\end{equation}

\smallskip
\noindent
{\it Draw 2, Step 4:} Draw the parameters for the GLM
for the continuum and absorption models. For this final step, we
condition only on $\bY\obs$, $\bY^B$, $\bY\updot$, and $\Ya\updot$, rather
than $\bY\aug$. We expect this substitution to improve the rate of
convergence of the sampler. Because the conditional
distribution is not from a standard family,  we use
a Metropolis Hastings step. In particular, we note that 
\begin{equation}
\mYa\updot_j
\Big|\btheta,\tilde{\blambda} \\ \dist \\ 
{\rm Poisson}
   \l(d_j\exp(\bX^A_j\btheta^A) \sum_{k=1}^K \tilde\lambda^k p_j^k \r)
 \ {\rm for} \  k\in\cK
\end{equation}

\begin{equation}
\mYa^C_j \Big|\btheta,\tilde{\blambda} \\ \dist \\
{\rm Poisson}\l(dE_j d_j\exp(\bX^C_j\btheta^C + \bX^A_j\btheta^A)\r)
\ {\rm for} \ j\in\cJ, 
\end{equation}
where $\tilde{\blambda}=(\tilde\lambda^1,\ldots,\tilde\lambda^K)$.
Since given $(\btheta^1,\ldots,\btheta^K)$ the log of each Poisson parameter
differs from 
a linear combination of $\btheta^C$ and $\btheta^A$ by a known constant, the
conditional posterior mode can easily be computed using a minor
modification of an Iteratively Reweighted Least Squares algorithm
\citep[e.g.,][]{this:88}. 
This algorithm can also account for the prior information described in
Section~3.2 and reports the curvature of the log posterior at the
mode. A multivariate t-distribution with four degrees of freedom with
the appropriate mode and (perhaps inflated) curvature can be used as a
jumping distribution to generate a proposal for the next
sample from the conditional distribution.
The relative mass of the jumping distribution
and actual conditional distribution of $\btheta^C$ and $\btheta^A$ at the
previous draw and proposed draw are
combined to determine if the proposal should be accepted or rejected
(in which case the previous draw is reused). Several (5 -- 10) proposals are
drawn at each iteration. We note that the same procedure can be used
to fit a GLM to the background counts (as was done in Section~4.2).

Although the MCMC methods detailed above may seem inhibiting as a
whole, each of the required steps are quite simple.  The power of
the MCMC methods described here (e.g., the Gibbs sampler) lies in
their ability to break complicated model fitting  tasks into a
succession of relatively simple tasks.  Our general strategy is to
use Bayes's theorem to derive a posterior distribution which hierarchically
accounts for the complexity both in the posited model and in data 
collection.  We then use modern statistical algorithms which devolve
model fitting into a sequence of relatively simple steps. We believe this
is a powerful strategy for dealing with the ever 
increasing power and sophistication of today's astronomical instruments.

\section{Internet Resources}

There are several internet sites where one can find papers describing
Bayesian methods and related software. The MCMC preprint service,

\noindent
{\tt http://www.mcs.surrey.ac.uk/Personal/S.Brooks/MCMC},

\noindent
and STATLIB, 

\noindent
{\tt http://lib.stat.cmu.edu} 

\noindent
are both large general statistical
sites that offer various software, preprints, and links that may
be of interest to astrophysicists. Three sites (that we know of) aim 
specifically at the interface of astrophysics and statistics:

\noindent
{\tt http://www.fas.harvard.edu/$\sim$vandyk/astrostat.html}

\noindent
 {\tt http://astrosun.tn.cornell.edu/staff/loredo/bayes}

\noindent
 {\tt http://www.astro.psu.edu/statcodes}

\clearpage

\if1\btex\bibliography{../../../TeXfiles/BibTeX/my,astro}\else

\fi

\clearpage

\if1\preprint
\begin{figure}[h]

\hskip -1.3in 
    \includegraphics[width=7in,height=3in]{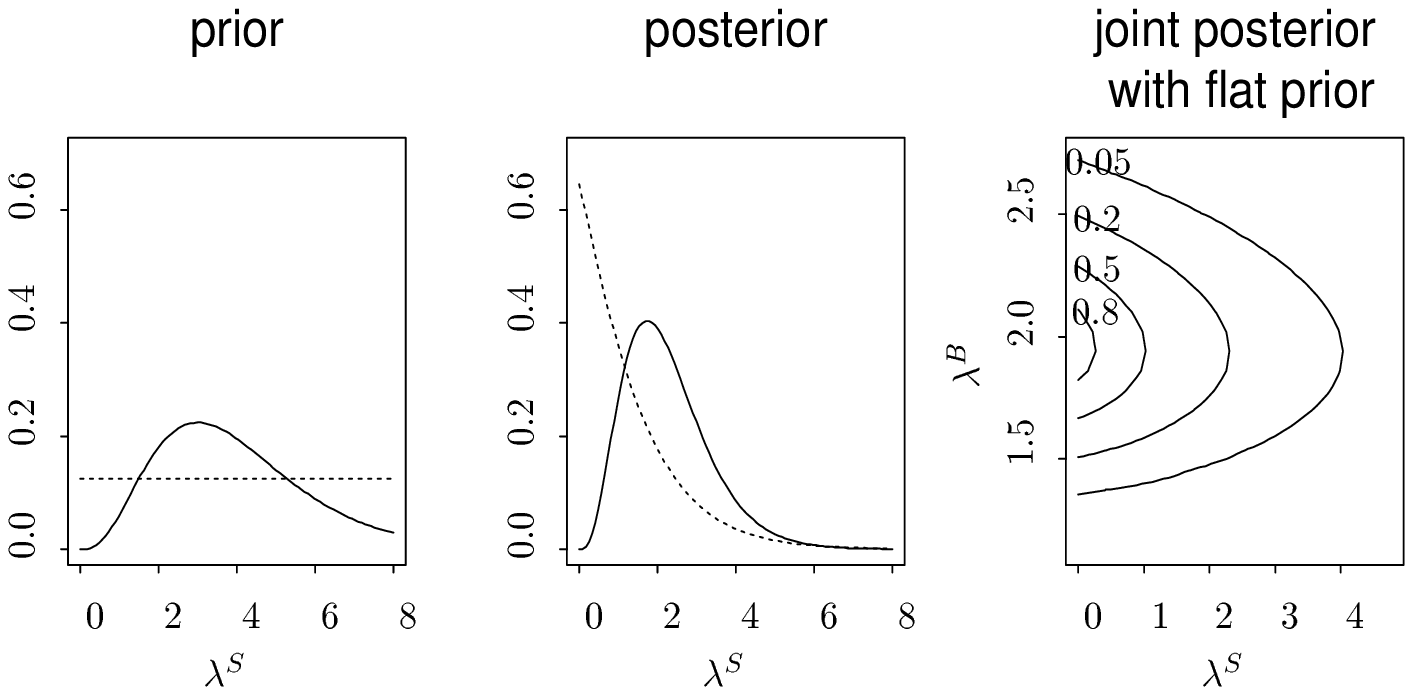}
    \caption{Combining information.  The figure illustrates the
     combination of the information contained in the data and  the
     prior into the posterior
     distribution.  The less informative dotted prior has less
     influence on its (dotted) posterior, which matches the low
     source count more closely than does the solid posterior.
     The joint posterior indicates the region of high posterior
     probability for both parameters under the non informative
     prior for $\lambda^S$.
     \label{fig:post_examp}}
\end{figure}

\begin{figure}[h]
\hskip -1in 
    \includegraphics[width=6.5in,height=4in]{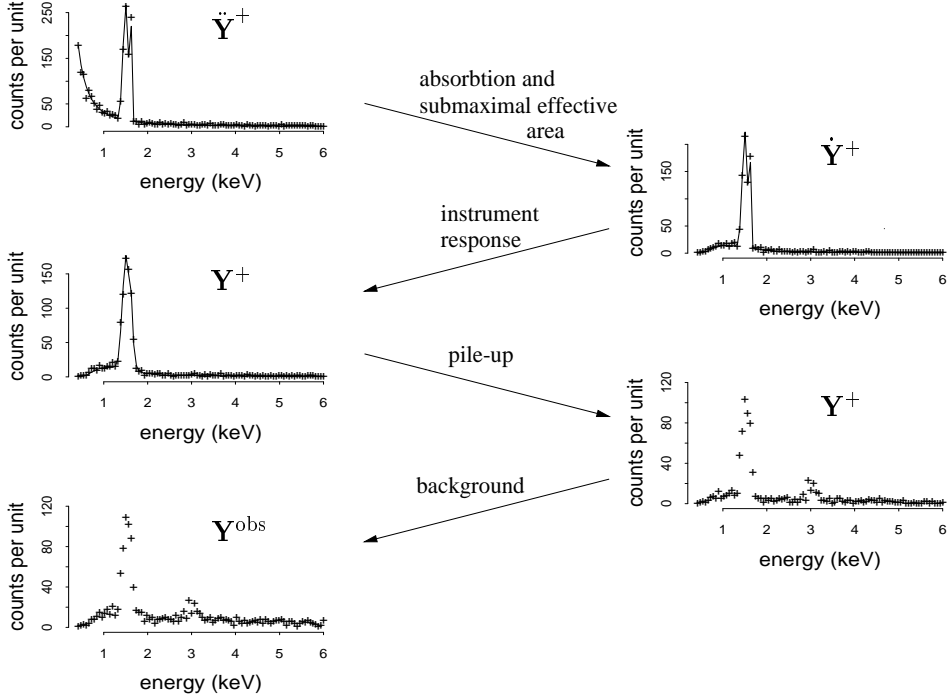}
    \caption{The Degradation of Counts.  The figure illustrates the various
      physical processes which significantly degrade the source model
      and result in the observed PHA counts. In particular, an
      artificial data set is used to illustrates (1) the
      absorption of (mostly low energy) counts, (2) the blurring of
      spectral features due to instrument response, (3) the shadows 
      caused by pile-up, and (4) the
      masking of features due to background. The solid lines represent
      the assumed model (in the first three plots) and the `$+$' sign
      the simulated data. The first plot illustrates the counts per
      maximum effective area per total exposure time per bin; the
      remaining plots illustrate  degraded counts per effective area per total
      exposure time per bin. Note that the effects of pile-up are
     included here for the sake of completeness; we do not deal with
      this aspect of the analysis in this paper. The symbols in the
      upper right of each plot are defined in Appendix~B.1.
\label{fig:datagen}}
\end{figure}


\begin{figure}[h]
  \begin{center}
    \if0\aastex
    \includegraphics[width=6in,height=6in]{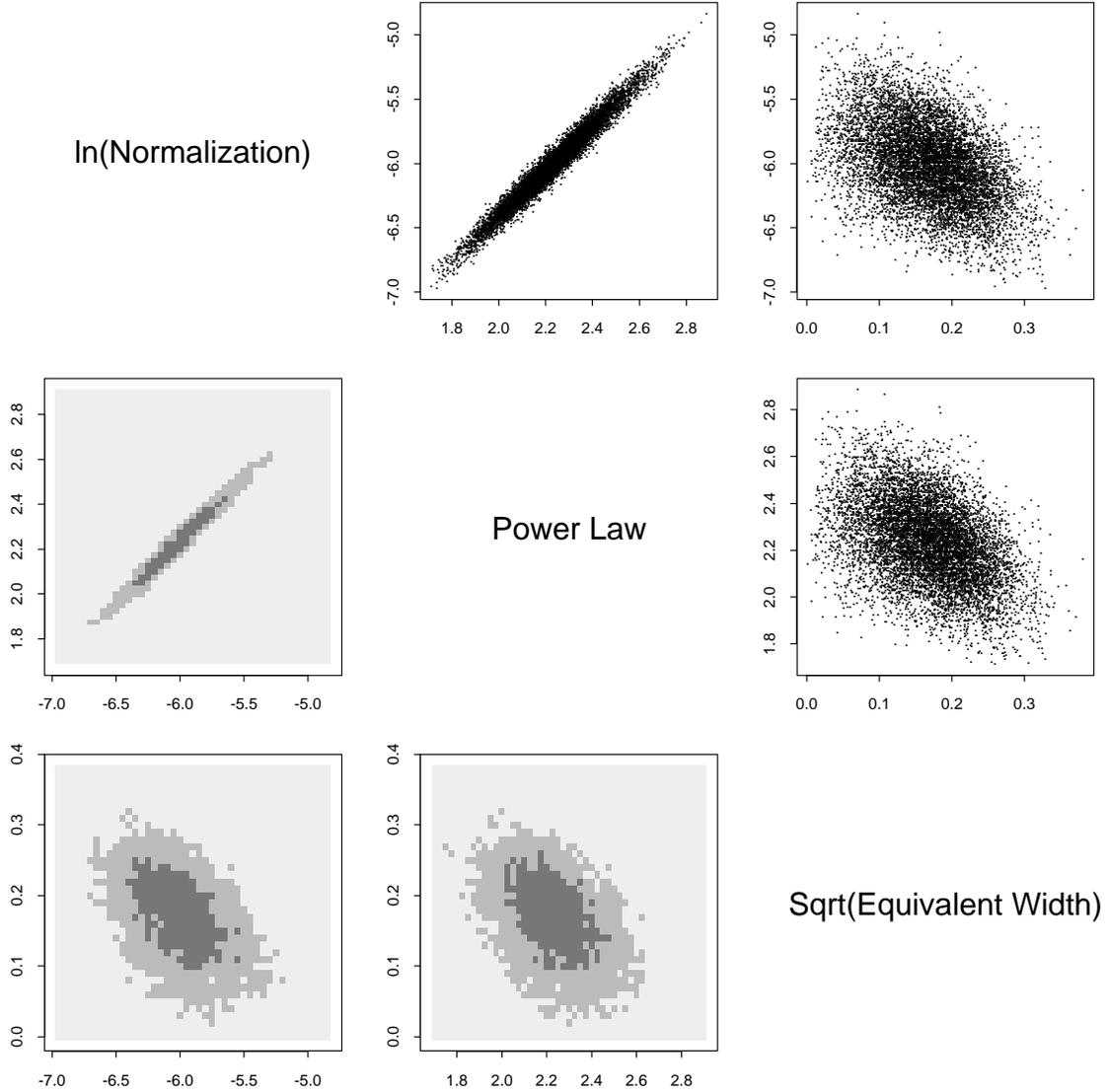}
    \else
    \includegraphics[width=6in,height=6in]{s0_postfigBIVAR.eps}
    \fi
    \caption{Posterior distributions of pairs of parameters obtained via
     MCMC. The plots show
      pairwise marginal posterior distributions for the model parameters in the
      analysis of Quasar S5 0014+813. The plots in the upper right are
      scatter plots of the Monte Carlo draws and indicate areas of
      highest posterior probability. The plots in the lower left are
      gray-scale images of the Monte Carlo approximations to 50\%
      (darker) and 90\% (lighter) credible regions. The text along the 
      diagonal labels the axes for each of the plots.
      \label{fig:quas-bipost}}
  \end{center}
\end{figure}

\begin{figure}[h]
  \begin{center}
    \if0\aastex
    \includegraphics[width=6in,height=6in]{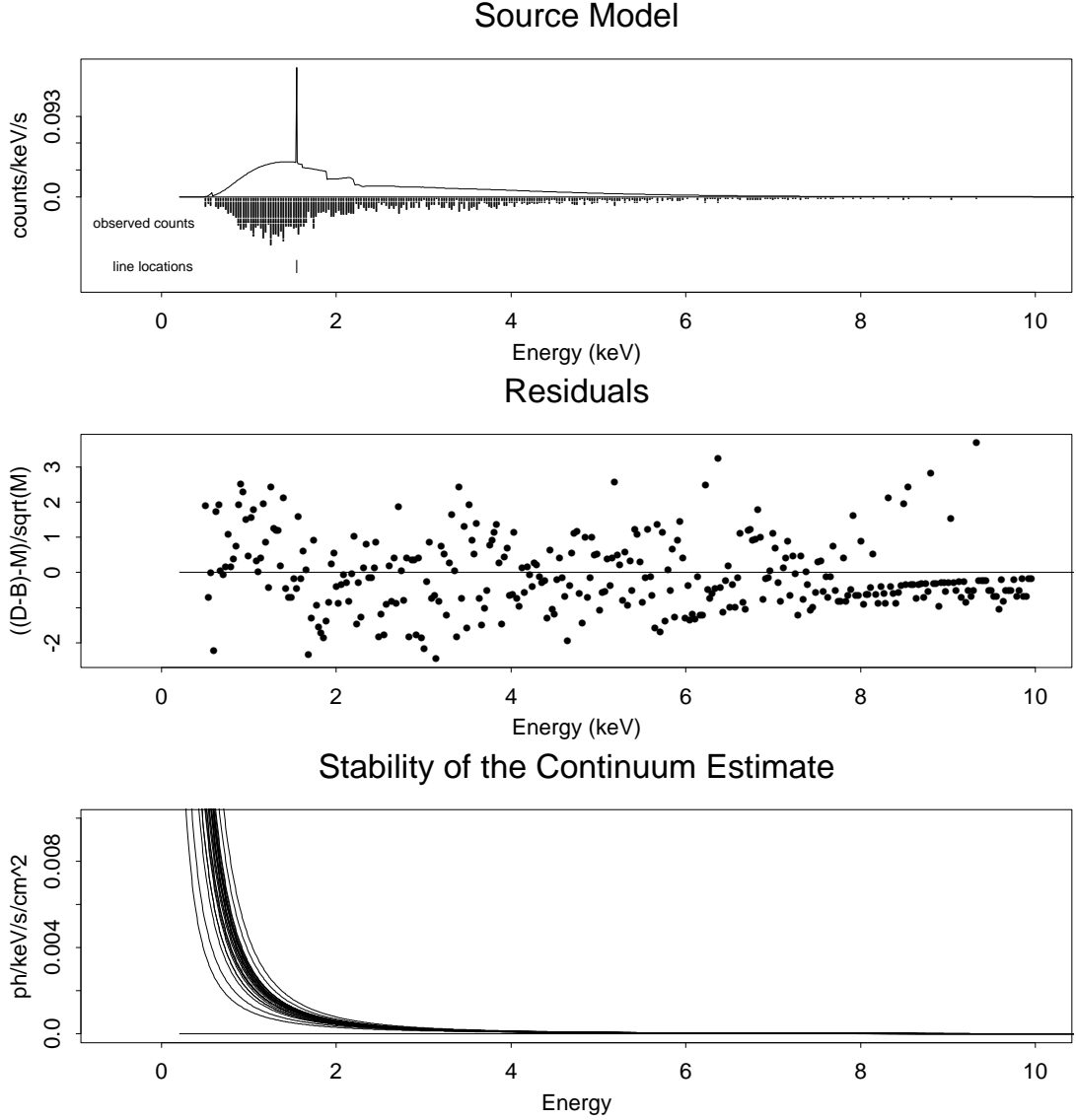}
    \else
    \includegraphics[width=6in,height=6in]{s0_stabilityfig.eps}
    \fi
    \caption{The Quasar S5 0014+813 model fit. This plot gives an
      overview of the fitted model. The first panel compares the
      fitted source model (corrected for effective area and
      absorption, but not the instrument's photon response matrix) with 
      the observed PHA counts. The second gives the
      residual for each PHA channel, which were computed
      by subtracting off background and standardizing by the model
      standard deviation. The final plot illustrates the
      stability of the continuum.
      \label{fig:quas-fit}}
  \end{center}
\end{figure}



\begin{figure}[h]
  \begin{center}
    \if0\aastex
    \includegraphics[width=6in,height=2in]{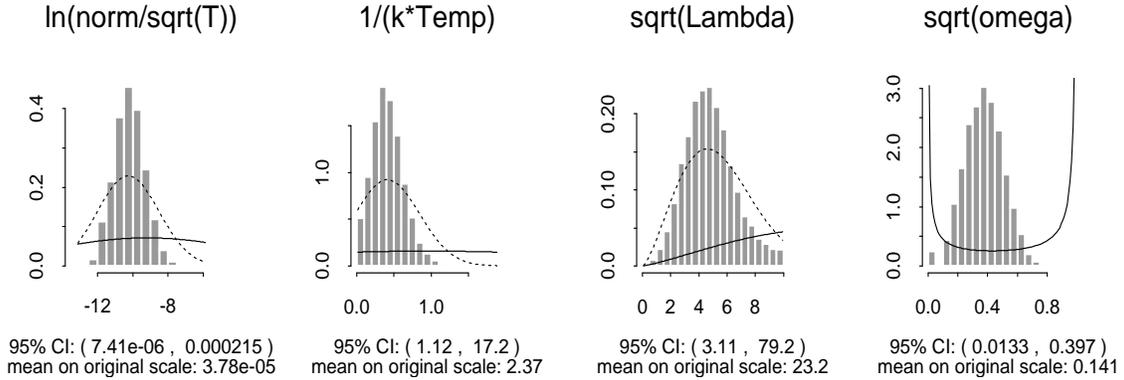}
    \else
    \includegraphics[width=6in,height=2in]{s0b1_postfig.eps}
    \fi
    \caption{Using ASCA/SIS to compute the prior. Here,
      Lambda is the expected model counts from lines and omega is the
      ratio of the total counts in the lines to the total counts in
      the spectrum (correcting for the effects of absorption and instrument
      response). Transformations of the parameters that produce the
      distribution nearest to the Gaussian are displayed. The listed
      means and credible intervals, however, refer to the original
      parameters. 
      The solid line in these
      plots represent the relatively diffuse priors used to compute
      the posterior distributions represented by the histograms based on the 
      SIS0 observation.  These posterior distributions 
      were in turn used to choose the  priors for the GIS data 
      after some
      dispersion was added, as represented by the dotted curves. We do
      not specify the prior for the proportion of source photons from
      the lines, $\Omega$, but rather this prior is implied by the
      other priors. The solid line is an approximation based on
      sampling from the prior and distributing the SIS0 counts to the
      continuum and lines after correcting for background.
      \label{fig:alpha-s0-post}}
  \end{center}
\end{figure}

\begin{figure}[h]
  \begin{center}
  \if0\aastex
  \includegraphics[width=6in,height=2in]{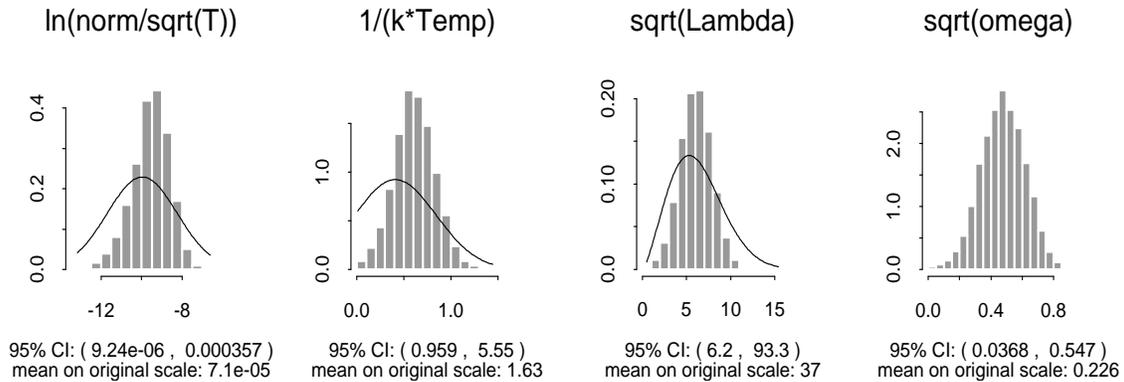}
  \else
  \includegraphics[width=6in,height=2in]{gis_postfig.eps}
  \fi
    \caption{Some marginal posterior distributions. Using the priors 
      computed with
      SIS0 data (solid lines) we fit the source model to the GIS
      data. The resulting marginal posterior distributions are illustrated here
      using normalizing transformations. The estimates and credible
      intervals are on the original scales.
      \label{fig:alpha-gis-post}}
  \end{center}
\end{figure}

\begin{figure}[h]
  \begin{center}
   \if0\aastex

\includegraphics[width=6in,height=6in]{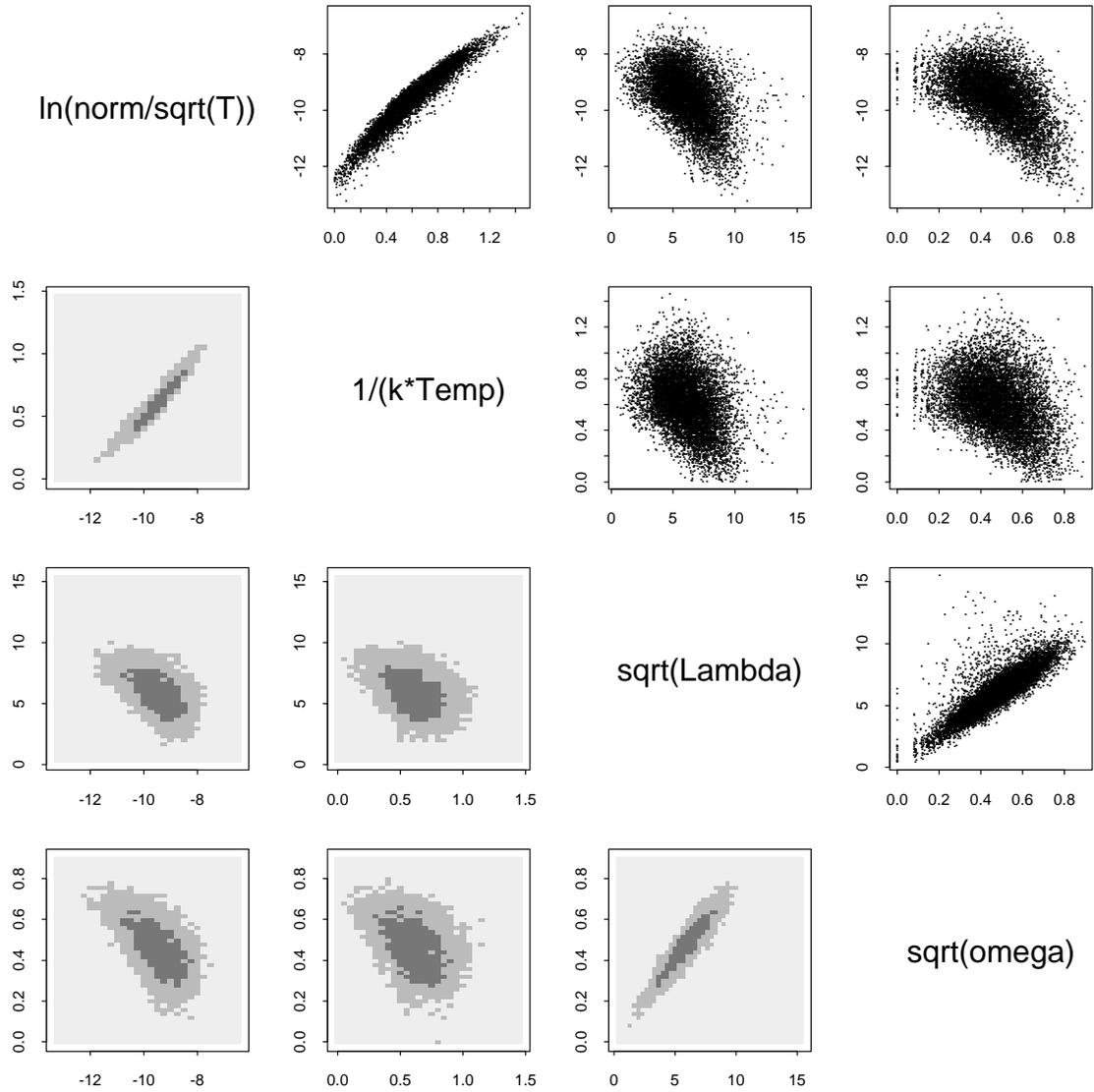}
   \else
   \includegraphics[width=6in,height=6in]{gis_postfigBIVARcont.eps}
   \fi
    \caption{Some bivariate marginal posterior distributions. These plots are 
      as
      described in Figure~\ref{fig:quas-bipost} and illustrate
      pairwise credible regions for the various model parameters.
      Again the text along the 
      diagonal labels the axes for each of the plots
      \label{fig:alpha-bipost}}
  \end{center}
\end{figure}

\begin{figure}[h]
  \begin{center}
   \if0\aastex
   \includegraphics[width=6in,height=6in]{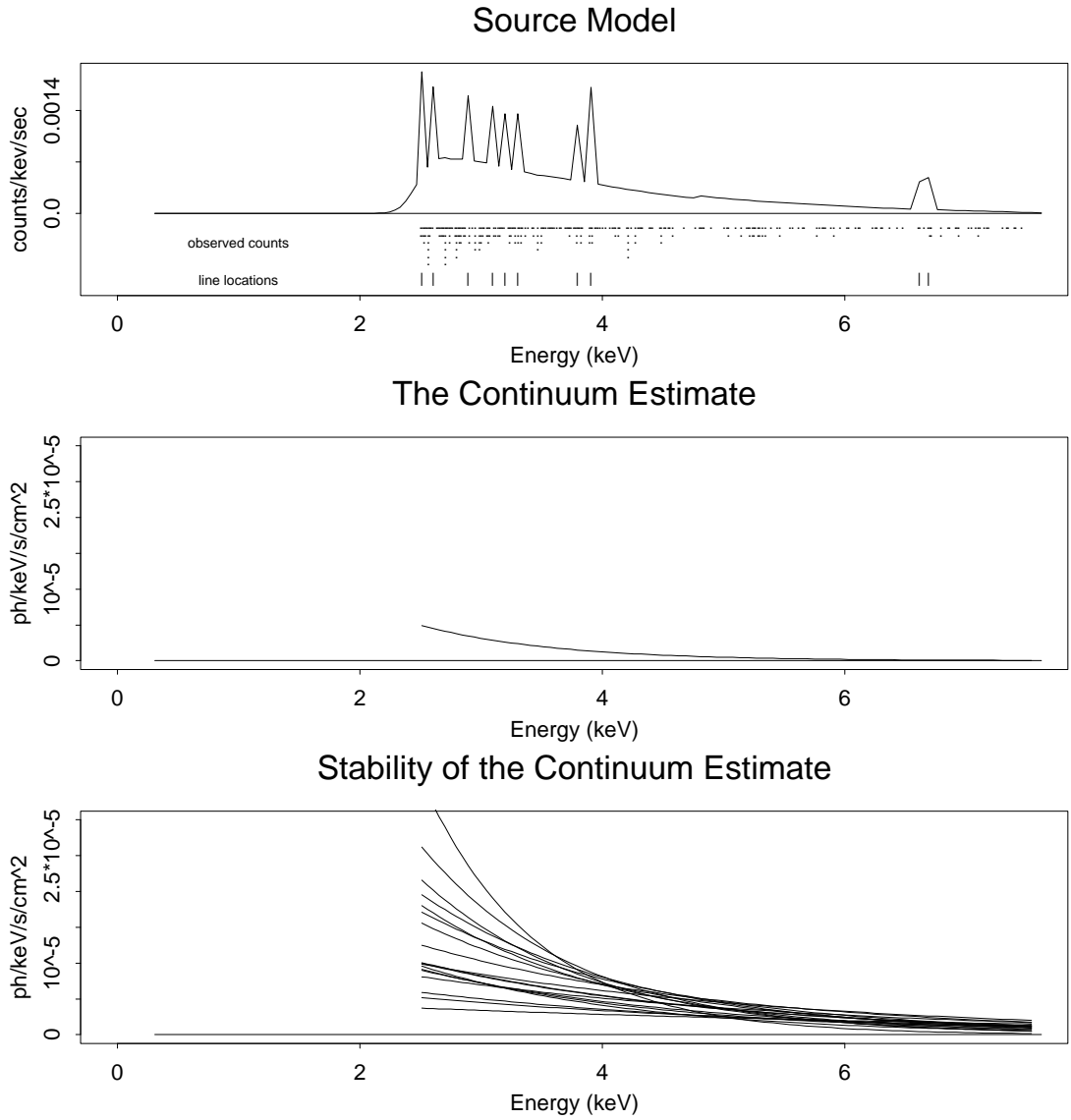}
   \else
   \includegraphics[width=6in,height=6in]{gis_stabilityfig.eps}
   \fi
    \caption{The model fit. These plots are as described in Figure~\ref{fig:quas-fit}
      -- note the the instability of the
      continuum due to the low counts.
      \label{fig:alpha-fit}}
  \end{center}
\end{figure}

\begin{figure}[h]
\hskip -1.3in
    \includegraphics[width=7in,height=2in]{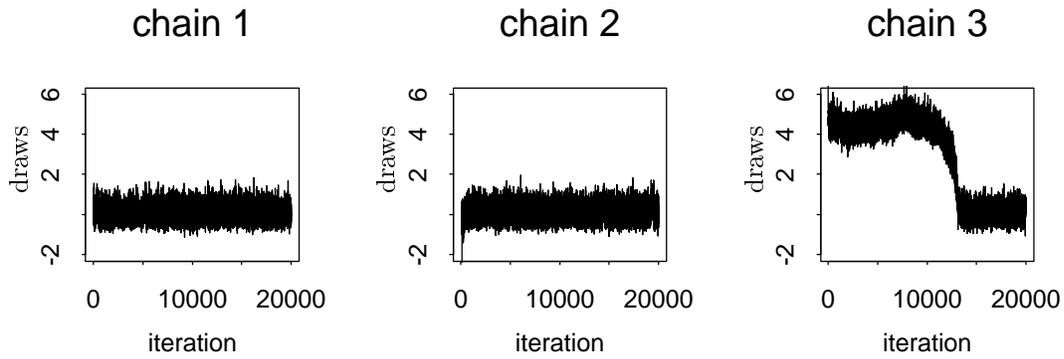}
    \caption{Several chains from a random-effects model.  Notice that
    chain 3 appears to have converged during the first 10000 iterations.
    Comparison with chain 1 and chain 2, however, makes it clear that
    chain 3 did not converge until after iteration 10000.\label{fig:slowconvg}}
\end{figure}

\begin{figure}[h]
\hskip -1.2in
    \includegraphics[width=7in,height=8in]{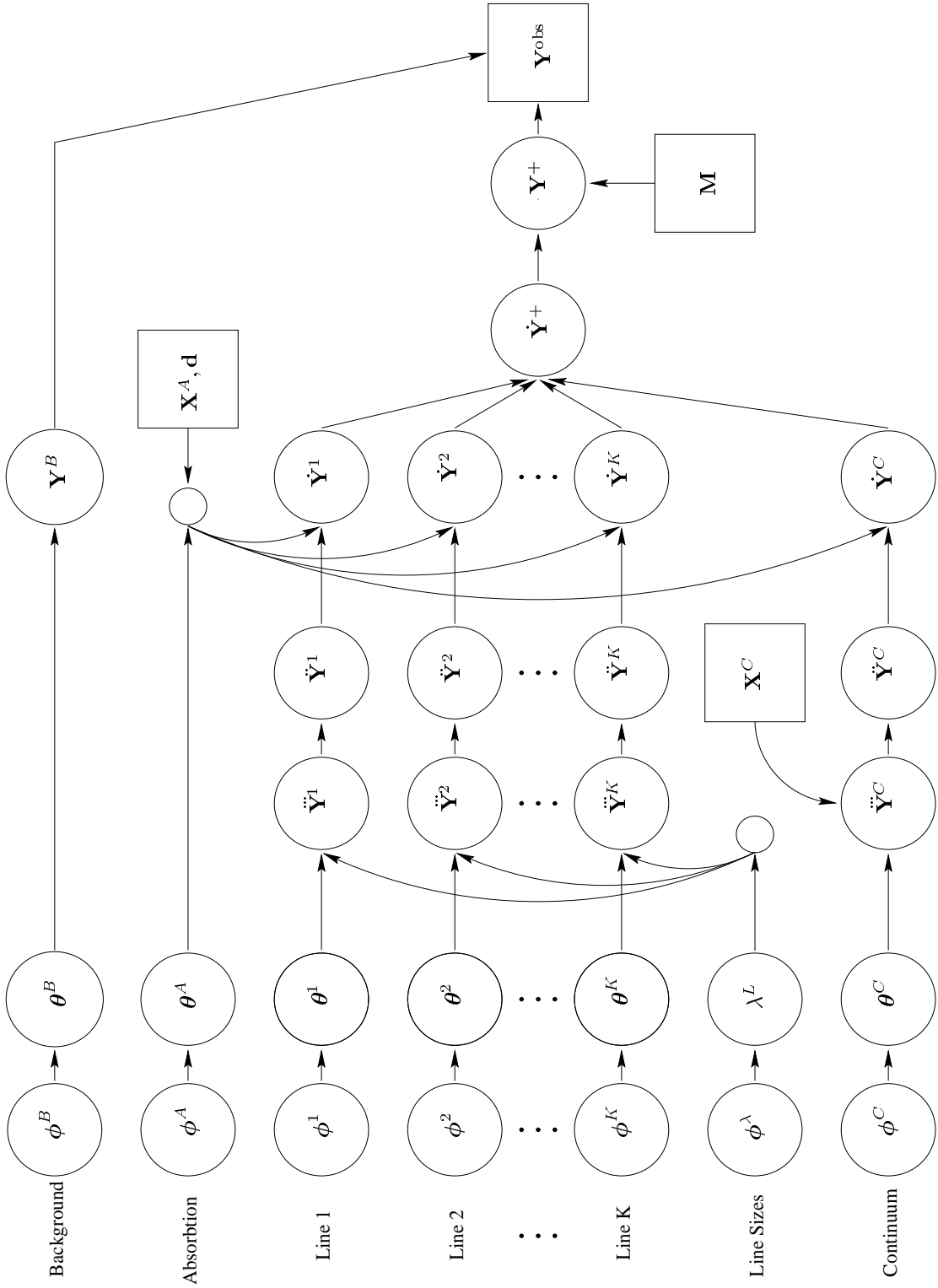}
    \caption{Caption appears on the following page.
    \label{fig:model}}
\end{figure}

\clearpage

\noindent 
    Caption for Figure 10:
    A graphical representation of the data-augmentation
    scheme. 
    Here $\bphi$ represents hyperparameters, $\btheta$ model parameters,
${{\bf Y}\kern -.65pc{}^{^{\textstyle\cdot\kern -3pt\cdot\kern -3pt\cdot}}}$
    true photon energies, $\Yb$ binned energies,
    $\Ya$ binned true photon energies after absorption accounting 
    for effective area, $\bY$ source counts in PHA channels, $\bY\obs$ the
    observed counts, $\bM$ the instrument response matrix, $\bd$ the effective
    area vector, $\bX^A$ and $\bX^C$ independent variables describing
    absorption and continuum respectively, circles represent  unobserved
    quantities, and squares observed quantities; details of the 
    subscripts and superscripts are given in the text. 
    The figure illustrates the interplay of the various model 
    parameters, hyperparameters, observed quantities, and data augmentation.
    As an example,
    the first arrow in the row labeled `background' corresponds to the 
    relationship between the background hyperparameters, $\bphi^B$, and 
    the background intensities, $\btheta^B$, e.g., 
    $\theta_l^B\dist\gamma(\phi_{l,1}^B,\phi_{l,2}^B)$. The second arrow
    corresponds to the Poisson nature of the background counts; see 
    Equation~\ref{eq:backgd}. The final background arrow illustrates 
    the background contamination of the observed PHA counts.

\fi

\if0\preprint
\begin{figure}[h]
    \caption{Combining information.  The figure illustrates the
     combination of the information contained in the data and  the
     prior into the posterior
     distribution.  The less informative dotted prior has less
     influence on its (dotted) posterior, which matches the low
     source count more closely than does the solid posterior.
     The joint posterior indicates the region of high posterior
     probability for both parameters under the non informative
     prior for $\lambda^S$.
     \label{fig:post_examp}}
\end{figure}

\begin{figure}[h]
    \caption{The Degradation of Counts.  The figure illustrates the various
      physical processes which significantly degrade the source model
      and result in the observed PHA counts. In particular, an
      artificial data set is used to illustrates (1) the
      absorption of (mostly low energy) counts, (2) the blurring of
      spectral features due to instrument response, (3) the shadows 
      caused by pile-up, and (4) the
      masking of features due to background. The solid lines represent
      the assumed model (in the first three plots) and the `$+$' sign
      the simulated data. The first plot illustrates the counts per
      maximum effective area per total exposure time per bin; the
      remaining plots illustrate  degraded counts per effective area per total
      exposure time per bin. Note that the effects of pile-up are
     included here for the sake of completeness; we do not deal with
      this aspect of the analysis in this paper. The symbols in the
      upper right of each plot are defined in Appendix~B.1.
\label{fig:datagen}}
\end{figure}

\begin{figure}[h]
    \caption{
    A graphical representation of the data-augmentation
    scheme. 
    Here $\bphi$ represents hyperparameters, $\btheta$ model parameters,
${{\bf Y}\kern -.82pc{}^{^{\textstyle\cdot\kern -3pt\cdot\kern -3pt\cdot}}}$       true photon
energies, $\Yb$ binned energies,
    $\Ya$ binned true photon energies after absorption accounting 
    for effective area, $\bY$ source counts in PHA channels, $\bY\obs$ the
    observed counts, $\bM$ the instrument response matrix, $\bd$ the effective
    area vector, $\bX^A$ and $\bX^C$ independent variables describing
    absorption and continuum respectively, circles represent  unobserved
    quantities, and squares observed quantities; details of the 
    subscripts and superscripts are given in the text. 
    The figure illustrates the interplay of the various model 
    parameters, hyperparameters, observed quantities, and data augmentation.
    As an example,
    the first arrow in the row labeled `background' corresponds to the 
    relationship between the background hyperparameters, $\bphi^B$, and 
    the background intensities, $\btheta^B$, e.g., 
    $\theta_l^B\dist\gamma(\phi_{l,1}^B,\phi_{l,2}^B)$. The second arrow
    corresponds to the Poisson nature of the background counts; see 
    Equation~\ref{eq:backgd}. The final background arrow illustrates 
    the background contamination of the observed PHA counts. 
    \label{fig:model}}
\end{figure}

\begin{figure}[h]
    \caption{
      Posterior distributions of pairs of parameters obtained via
      MCMC. The plots show
      pairwise marginal posterior distributions for the model parameters in the
      analysis of Quasar S5 0014+813. The plots in the upper right are
      scatter plots of the Monte Carlo draws and indicate areas of
      highest posterior probability. The plots in the lower left are
      gray-scale images of the Monte Carlo approximations to 50\%
      (darker) and 90\% (lighter) credible regions. The text along the 
      diagonal labels the axes for each of the plots.
      \label{fig:quas-bipost}}
\end{figure}

\begin{figure}[h]
    \caption{The Quasar S5 0014+813 model fit. This plot gives an
      overview of the fitted model. The first panel compares the
      fitted source model (corrected for effective area and
      absorption, but not the instrument's photon response matrix) with 
      the observed PHA counts. The second gives the
      residual for each PHA channel, which were computed
      by subtracting off background and standardizing by the model
      standard deviation. The final plot illustrates the
      stability of the continuum.
      \label{fig:quas-fit}}
\end{figure}

\begin{figure}[h]
    \caption{Using ASCA/SIS to compute the prior. Here,
      Lambda is the expected model counts from lines and omega is the
      ratio of the total counts in the lines to the total counts in
      the spectrum (correcting for the effects of absorption and instrument
      response). Transformations of the parameters that produce the
      distribution nearest to the Gaussian are displayed. The listed
      means and credible intervals, however, refer to the original
      parameters. 
      The solid line in these
      plots represent the relatively diffuse priors used to compute
      the posterior distributions represented by the histograms based on 
      the SIS0
      observation.  These posterior distributions 
      were in turn used to choose the  priors for the GIS data 
      after some
      dispersion was added, as represented by the dotted curves. We do
      not specify the prior for the proportion of source photons from
      the lines, $\Omega$, but rather this prior is implied by the
      other priors. The solid line is an approximation based on
      sampling from the prior and distributing the SIS0 counts to the
      continuum and lines after correcting for background.
      \label{fig:alpha-s0-post}}
\end{figure}

\begin{figure}[h]
    \caption{Some marginal posterior distributions. Using the priors 
      computed with
      SIS0 data (solid lines) we fit the source model to the GIS
      data. The resulting marginal posterior distributions are illustrated 
      here
      using normalizing transformations. The estimates and credible
      intervals are on the original scales.
      \label{fig:alpha-gis-post}}
 \end{figure}

\begin{figure}[h]
    \caption{Some bivariate marginal posterior distributions. These plots 
      are as
      described in Figure~\ref{fig:quas-bipost} and illustrate
      pairwise credible regions for the various model parameters.
      Again the text along the 
      diagonal labels the axes for each of the plots.
      \label{fig:alpha-bipost}}
\end{figure}

\begin{figure}[h]
    \caption{The model fit. These plots are as described in 
      Figure~\ref{fig:quas-fit} -- note the the instability of the
      continuum due to the low counts.
      \label{fig:alpha-fit}}
\end{figure}

\begin{figure}[h]
    \caption{Several chains from a random-effects model.  Notice that
    chain 3 appears to have converged during the first 10000 iterations.
    Comparison with chain 1 and chain 2, however, makes it clear that
    chain 3 did not converge until after iteration 10000.\label{fig:slowconvg}}
\end{figure}
\fi

\clearpage

\begin{table}
\begin{center}
\begin{tabular}{ll}
\hline\hline
Terms & Defined and/or Discussed \\[0.5ex]
\hline
absorption & Sections~3.1, 3.4 \\
Bayes's Theorem & Equation~\ref{eq:bayesthm} \\
conjugate prior distribution & Footnote 10 \\
credible interval & Section~2.1 \\
data augmentation & Sections~1, 2.3, B.1 \\
data augmentation algorithm & Sections~2.3, A.1 \\
effective area & Footnote~3 \\
equivalent width & Footnote 12 \\
gamma distribution & Footnote 7 \\
GLM or generalized linear model & Footnote~4, Section~3.2 \\
Gibbs sampler & Section~2.3, A.3 \\
hierarchical model & Footnote 5 \\
hyperparameters & Section 2.1 \\
improper distribution & Footnote 8 \\
MCMC or Markov chain Monte Carlo & Sections 2.3, A.1, A.2, A.3, A.4 \\
marginal distribution & Section~2.1, Equation~\ref{eq:marg} \\
maximum effective area & Footnote 9 \\
model & Section 1 \\
Monte Carlo integration & Section 2.2 \\
multinomial distribution & Footnote 16 \\
non-informative prior distribution & Section 2.1 \\
nuisance parameter & Section 2.1, Equation~\ref{eq:marg} \\
observed-data model & Section 1 \\
Poisson distribution & Footnote 2 \\
posterior distribution & Equation~\ref{eq:bayesthm} \\
prior distribution & Equation~\ref{eq:bayesthm} \\
PHA or pulse height amplitude & Footnote 1 \\
source model & Section 1 \\
\hline
\hline
\end{tabular}\\[0.5ex]
\end{center}
\caption{Index of various terms discussed and defined in the paper.
\label{tbl:index}}
\end{table}

\clearpage

\begin{table}
\begin{center}
\begin{tabular}{ccc}
\hline\hline
Parameter & Estimate &95\% Interval\\[0.5ex]
\hline
Power Law        & 2.23      & (1.90, 2.58) \\
Normalization    &2.47e-3    & (1.37e-3, 4.55e-3) \\
Absorption       &2.05       & (1.43, 2.71) \\
Equivalent Width & 0.0282 keV & (0.0023, 0.0788) keV\\
\hline
\hline
\end{tabular}\\[0.5ex]
\end{center}
\caption{Fitted values and credible regions for the Quasar data.\label{tbl:quasar}}
\end{table}

\clearpage

\begin{table}
\begin{center}
\begin{tabular}{lcl}
\hline\hline
Variable & Notation & Range\\[0.5ex]
\hline
The photon energy&$\mYc_i$& Positive, measured in keV\\
Indicator for Background&$Z_i^B$&1 for background photons\\
&&0 for other photons\\
Indicator for Continuum&$Z_i^C$&1 for continuum photons\\
&&0 for other photons\\
Indicator for Line $k$, &$Z_i^k$&1 for photons from line $k$\\
$ \qquad $ for $k=1,\ldots K$&&0 for other photons\\
Indicator for Absorption &$Z_i^A$&1 for absorbed photons\\
&&0 for other photons\\
\hline
\hline
\end{tabular}\\[0.5ex]
\end{center}
\caption{Variables associated with each photon, for
$i=1,\ldots,N$.
\label{tbl:data}}
\end{table}

\clearpage

\begin{table}
\begin{center}
\begin{tabular}{lcl}
\hline\hline
Variable & Notation & Range\\[0.5ex]
\hline
The unbinned energies&$\Yc^s$& Positive, measured in keV, $s\in\cS$\\
The binned energies &$\Yb^s_j$& Counts for $j\in\cJ, s\in\cS$ \\
The binned energies &$\Ya_j^s$& Counts for $j\in\cJ, s\in\cS$ \\
 \  after absorption & & \\
The blurred PHA counts&$\bY\updot_l$& Counts for $l \in \cL$\\
 \  without background & &\\
The observed data&$\bY\obs_l$& Counts for $l \in \cL$\\
\hline
\hline
\end{tabular}\\[0.5ex]
\caption{Summary statistics for the spectral model.
\label{tbl:data_summaries}}
\end{center}
\end{table}

\end{document}